\documentclass[apj]{emulateapj}
\usepackage{graphicx}
\usepackage[toc,page]{appendix}
\usepackage{natbib}
\usepackage{color}
\usepackage{threeparttable}
\usepackage{hyperref}
\usepackage{breakurl}
\begin{document}

%\slugcomment{Draft version: \today}
\slugcomment{Accepted for publication in the Astrophysical Journal}
\title{The NEWFIRM Medium-band Survey: Photometric Catalogs, Redshifts and the Bimodal Color Distribution of Galaxies out to $z\sim3$}
\email{katherine.whitaker@yale.edu}
\author{Katherine E. Whitaker\altaffilmark{1,2}, Ivo Labb\'{e}\altaffilmark{3,2},
Pieter G. van Dokkum\altaffilmark{1,2}, Gabriel Brammer\altaffilmark{1,4,2}, Mariska Kriek\altaffilmark{5,6,2}, 
Danilo Marchesini\altaffilmark{7,2}, Ryan F. Quadri\altaffilmark{8,2}, Marijn Franx\altaffilmark{8},
Adam Muzzin\altaffilmark{1,2}, Rik J. Williams\altaffilmark{3,2}, Rachel Bezanson\altaffilmark{1}, 
Garth D. Illingworth\altaffilmark{9}, Kyoung-Soo Lee\altaffilmark{1,2}, Britt Lundgren\altaffilmark{1},
Erica J. Nelson\altaffilmark{1}, Gregory Rudnick\altaffilmark{10,2}, Tomer Tal\altaffilmark{1}, 
David A. Wake\altaffilmark{1}}

\altaffiltext{1}{Department of Astronomy, Yale University, New Haven, CT 06511}
\altaffiltext{2}{Visiting Astronomer, Kitt Peak National Observatory, National Optical
Astronomy Observatory, which is operated by the Associations of Univerities for Research
in Astronomy (AURA) under cooperative agreement with the National Science Foundation.}
\altaffiltext{3}{Carnegie Observatories, Pasadena, CA 91101}
\altaffiltext{4}{European Southern Observatory, Alonso de C\'{o}rdova 3107, Casilla 19001, Vitacura, Santiago, Chile}
\altaffiltext{5}{Department of Astrophysical Sciences, Princeton University, Princeton, NJ 08544}
\altaffiltext{6}{Harvard-Smithsonian Center for Astrophysics, 60 Garden Street, Cambridge, MA 02138, USA}
\altaffiltext{7}{Department of Physics and Astronomy, Tufts University, Medford, MA 02155}
\altaffiltext{8}{Sterrewacht Leiden, Leiden University, NL-2300 RA Leiden, The Netherlands}
\altaffiltext{9}{UCO/Lick Observatory, Pasadena, CA 91101}
\altaffiltext{10}{Department of Physics and Astronomy, University of Kansas, Lawrence, KS 66045}

\shortauthors{Whitaker et al.}
\shorttitle{Photometric Catalogs and Redshifts from the NMBS}

\begin{abstract}
We present deep near-infrared (NIR) medium-bandwidth photometry over the wavelength range 1--1.8$\mu$m
in the All-wavelength Extended Groth strip International Survey (AEGIS) and Cosmic Evolution Survey (COSMOS)
fields.  The observations were carried out as part of the NEWFIRM Medium-Band Survey (NMBS),  
an NOAO survey program on the Mayall 4m 
telescope on Kitt Peak using the NOAO Extremely Wide-Field Infrared Imager (NEWFIRM). 
In this paper, we describe the full details of the observations, data reduction and photometry
for the survey.  We also present a public $K$-selected photometric catalog, along with accurate photometric redshifts.  
The redshifts are computed with 37 (20) filters in the COSMOS (AEGIS) fields, combining the NIR medium-bandwidth
data with existing ultraviolet (UV; {\it Galaxy Evolution Explorer}), visible and NIR (Canada--France--Hawaii Telescope 
and {\it Subaru}) and mid-IR ({\it Spitzer}/IRAC) imaging.  We find excellent agreement with publicly available spectroscopic 
redshifts, with $\sigma_{z}/(1+z)\sim1$--2\% for $\sim$4000 galaxies at $z=0$--3.  The
NMBS catalogs contain $\sim13,000$ galaxies at $z>1.5$ with accurate photometric 
redshifts and rest-frame colors.  Due to the increased spectral resolution obtained with the five NIR medium-band filters, 
the median 68\% confidence intervals of the photometric 
redshifts of both quiescent and star-forming galaxies are a factor of $\sim2$ 
times smaller when comparing catalogs with medium-band NIR photometry to NIR broadband photometry. 
%Additionally, the fraction of catastrophic outliers $>5\sigma$ in the photometric-spectroscopic redshift comparison
%increases by a factor of four and $\sigma_{\mathrm{z}}/(1+z)$ increases to $\sim5$\%
%when comparing the NMBS catalogs to mock broadband catalogs.
We show evidence for a clear bimodal color distribution between quiescent and star-forming galaxies that persists 
to $z\sim3$, a higher redshift than has been probed so far.
\end{abstract}


\section{Introduction}
\label{sec:intro}

The growing number of deep, wide-field multiwavelength surveys over multiple bands in the past decade
has led to new insights into the physical processes that govern galaxy formation
and evolution.  With the advent of larger ground-based telescopes with improved 
NIR cameras and a new generation of space-based telescopes, surveys have
pushed to higher redshifts and further down the luminosity and mass functions.  
Although broadband photometric surveys of statistically significant samples of galaxies now span
large portions of cosmic time, there exists a fundamental limitation: the 
accuracy of the photometric redshift estimates.  
Accurate photometric redshifts are necessary to measure reliable rest-frame colors,
stellar population properties, and the environmental densities.

Deep optical medium-band photometric
surveys such as the Classifying Object by Medium-Band Observations-17~\citep[COMBO-17;][]{Wolf03} and
the Cosmic Evolution Survey (COSMOS) with 30-bands~\citep{Ilbert09}
have photometric redshifts accuracies of $\sigma_{\Delta z/(1+z_{\mathrm{spec}})}\sim1$\% for large samples of galaxies
out to $z\sim1$.  By using medium-bandwidth filters, these surveys have improved the accuracy of
photometric redshifts by factors of 2--3.

As prominent rest-frame optical spectral features -- notably the 4000\AA\ break-- 
are shifted into the near-infrared at $z>1.5$, most
studies at high redshift have focused on those galaxies that are relatively bright at observed optical (rest-frame
ultraviolet) wavelengths \citep[e.g.,][]{Steidel96, Steidel99}.  Consequently, these studies have missed relatively
red galaxies~\citep[see, e.g.,][]{Franx03,vanDokkum06}.  Concessions must be made for studies of complete samples of 
high redshift galaxies, where one can either work with small, bright spectroscopic samples~\citep{Cimatti02, Kriek08a},
or use large samples that rely on less accurate broadband photometric 
redshifts~\citep[e.g.][and many other studies]{Dickinson03, Fontana06}.  

%=== Fig 1
\begin{figure*}[t!]
\leavevmode
\centering
\includegraphics[scale=0.6]{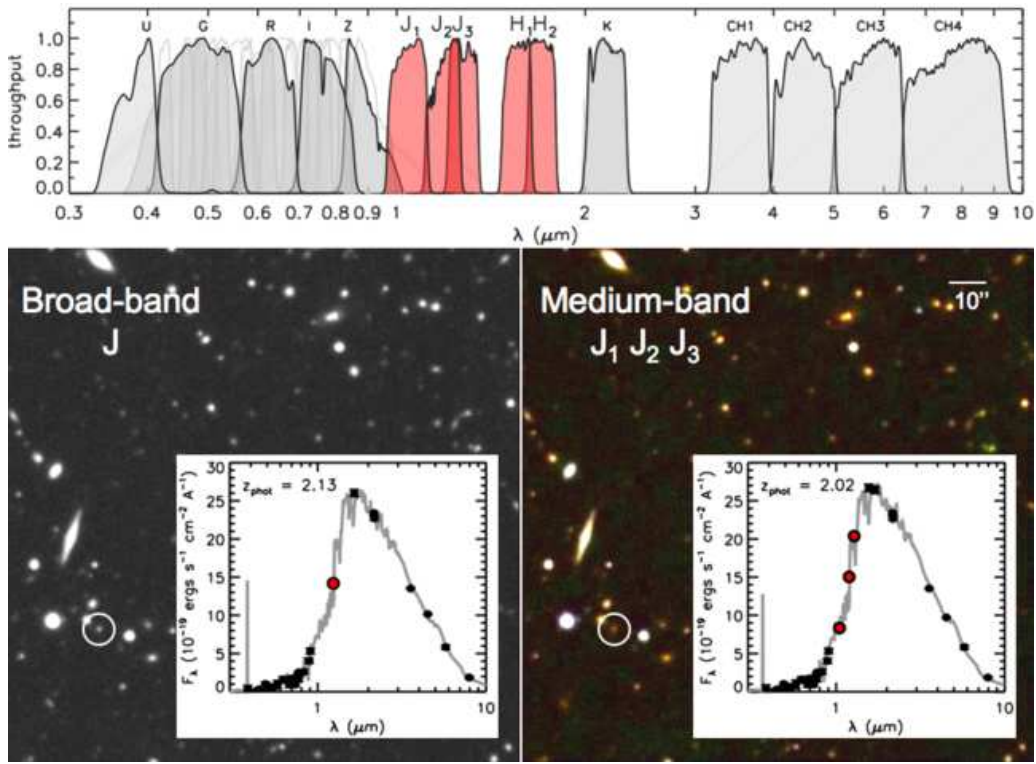}
\caption{A small region of the COSMOS field (roughly 2' on a side), shown for a broadband $J$-filter (left)
and a three color image from the medium-band $J_{1}J_{2}J_{3}$ filters (right).  The red object in the
bottom left of the image highlights the spectral features that we are able to resolve within only 0.2$\mu$m in
wavelength (see inset SED).
The $J$-filters trace out the Balmer/4000\AA\ break with higher resolution than the standard broadband filter,
allowing for an accurate photometric redshift of this quiescent galaxy at $z$=2.02.
The normalized transmission curves of the filters in the NMBS catalogs are shown in the top panel; the additional 
COSMOS filters (e.g. the {\it Subaru} optical medium-band filters) are shown in grey-scale behind the five optical filters. }
\label{fig:J123}
\end{figure*}

Here, following previous medium-band surveys in the optical, 
we extend the redshift range of accurate photometric redshifts out to $z\sim3$ using medium-band NIR filters. 
The NEWFIRM Medium-Band Survey (NMBS) is a 75 night NOAO survey program on the Kitt Peak 4m telescope.

An outline of the paper follows.
%In this work, we derive a multiband $K$-selected galaxy catalog from the NMBS data, 
%calculating the photometric redshifts and rest-frame colors for all objects.  
We introduce the details of the NMBS in \S~\ref{sec:survey} and describe the image
processing and optimization in \S~\ref{sec:reduction}.  The source detection and  
photometry are next described in \S~\ref{sec:source}, elaborating on all of the details 
involved in  extracting the NMBS catalogs.  In \S~\ref{sec:othercatalogs}, we compare the NMBS catalogs
to other publicly available catalogs within the fields, finding good agreement.  We derive photometric 
redshifts and rest-frame colors in \S~\ref{sec:photoz} and show that the accurate
redshifts and colors enable us to identify quiescent galaxies out to $z\sim3$ in \S~\ref{sec:quiescent}.
Finally, we demonstrate the improvements enabled by the medium-bandwidth filters relative 
to the standard broadband filters by comparing the confidence intervals of the photometric
redshifts in \S~\ref{sec:benefits}.  
A summary of the survey can be found in \S~\ref{sec:summary}.

We assume a $\Lambda$CDM cosmology with $\Omega_{M}$=0.3, $\Omega_{\Lambda}$=0.7, 
and $H_{0}$=70 km s$^{-1}$ Mpc$^{-1}$ throughout the paper.  All magnitudes are 
given in the AB system.

\section{The NEWFIRM Medium-Band Survey}
\label{sec:survey}

The NMBS employs a new technique of using medium-bandwidth NIR filters to sample the
Balmer/4000\AA\ break from $1.5<z<3.5$ at a higher resolution than the standard
broadband NIR filters~\citep{vanDokkum09a}, thereby improving the accuracy of photometric redshifts.
A custom set of five medium bandwidth filters in the wavelength range of
1--1.8$\mu$m were fabricated for the NEWFIRM camera on the Mayall
4m telescope on Kitt Peak for the NMBS.  The $J_{1}$-band is similar to the $Y$-band,
the canonical $J$-band is split into two filters $J_{2}$ and $J_{3}$, and the $H$-band is split into
two filters $H_{1}$ and $H_{2}$ (see the top panel in Figure~\ref{fig:J123}).  
The full details of the medium-band filters can be found in \citet{vanDokkum09a}.

Figure~\ref{fig:J123} demonstrates the power of the medium-band filters, comparing
a traditional $J$-band image to a three-color image comprised of the $J_{1}J_{2}J_{3}$ filters.
Although the wavelength range of the three medium bandwidth $J$-filters only covers 0.2$\mu$m,
the filters are able to resolve strong spectral features such as Balmer/4000\AA\ breaks or emission 
lines.  For example, $J_{1}J_{2}J_{3}$ trace out the 4000\AA\ break of the massive, quiescent
galaxies at $z=2.02$ shown in Figure~\ref{fig:J123}.  When using the broadband $J$ and $H$ filters alone,
the photometric redshift uncertainty is $\sim$4\% in $(z_{\mathrm{NMBS}}-z_{\mathrm{broadband}})$/($1+z_{\mathrm{NMBS}}$), typical
of the inherent uncertainties associated with broadband photometric redshifts.  Through sampling the 
Balmer/4000\AA\ break region of the spectral energy distributions (SED) of the galaxies at $1.5<z<3.5$
with higher resolution, the uncertainties in the photometric redshifts and rest-frame colors 
decreases by about a factor of two (see \S~\ref{sec:benefits}).  
 
\subsection{Field Selection}
\label{sec:field}

The survey targets two fields within the COSMOS~\citep{Scoville07} and AEGIS~\citep{Davis07} surveys, 
chosen to take advantage of the wealth of publicly-available ancillary data over a broad wavelength range 
(see \S~\ref{sec:ancillary}).
The NEWFIRM 27$^{\prime}$.6$\times$27$^{\prime}$.6 pointing within the COSMOS field is centered at 
$\alpha$ = 9$^{h}$59$^{m}$53.3$^{s}$, 
$\delta$ = +02$^{\circ}$24$^{m}$08$^{s}$ (J2000).  This pointing overlaps with the $z$COSMOS deep
redshift survey \citep{Lilly07} and the upcoming Ultra-VISTA 
survey\footnote{\url{http://www.eso.org.sci/observing/policies/PublicSurveys/}}.
The pointing in the AEGIS field is centered at $\alpha$ = 14$^{h}$18$^{m}$00$^{s}$, 
$\delta$ = +52$^{\circ}$36$^{m}$07$^{s}$ (J2000).  

\subsection{Observations}
\label{sec:observations}

The observations were carried out on the Mayall 4m telescope on Kitt Peak using the 
NEWFIRM camera with the $J_{1}J_{2}J_{3}H_{1}H_{2}$ medium bandwidth filters and the $K$ broadband filter.
Data were taken over the 
2008A, 2008B and 2009A semesters, for a total of 75 nights.
The two main fields were observed whenever conditions were reasonable (i.e., no significant cirrus and seeing typically 
$<$1.5$^{\prime\prime}$) and they were available at an airmass $<$2 and $>$20$^{\circ}$ away
from the moon.

\begin{table*}[t!]
  \caption{Summary of Observations}
  \centering
  \begin{threeparttable}
    \begin{tabular}{lccccccc}
      \hline
      \hline
      Field~~~~~~~~~~~~~~~~~~~~&  Filter  & On-Sky Time\tnote{a} [hr] & Integration Time\tnote{b} [hr] & Image Quality\tnote{c} [$^{\prime\prime}$] & 5$\sigma$ Depth\tnote{d} & Aperture Correction\tnote{e} & Zero Point \\
%      (1) & (2) & (3) & (4) & (5) & (6) & (7) & (8) \\
      \hline
      AEGIS\dotfill  & J1 & 29.3 & 25.0 & 1.13 & 25.2 & 1.9 & 23.31 \\ 
      & J2 & 25.9 & 24.1 & 1.16 & 25.3 & 1.9 & 23.35 \\
      & J3 & 29.1 & 22.9 & 1.08 & 24.5 & 1.8 & 23.37 \\
      & H1 & 26.5 & 22.3 & 1.10 & 24.1 & 2.0 & 23.59 \\
      & H2 & 20.8 & 17.1 & 1.06 & 24.4 & 1.8 & 23.61 \\  
      & K  & 22.2 & 12.8 & 1.08 & 24.2 & 1.9 & 23.85 \\
      \hline
      COSMOS\dotfill & J1 & 31.2 & 24.9 & 1.19 & 25.1 & 2.0 & 23.31 \\
      & J2 & 21.9 & 19.6 & 1.17 & 24.8 & 1.9 & 23.35 \\
      & J3 & 27.5 & 24.8 & 1.12 & 24.7 & 1.9 & 23.37 \\
      & H1 & 22.3 & 14.7 & 1.03 & 24.2 & 1.8 & 23.59 \\
      & H2 & 20.0 & 17.0 & 1.24 & 24.0 & 2.1 & 23.61 \\
      & K  & 17.0 & 11.8 & 1.08 & 24.2 & 1.9 & 23.85 \\
      \hline
    \end{tabular}
    \begin{tablenotes}
    \item[a] The total on-sky time represents all usable data, including aborted sequences and frames
      outside of the weather specifications or seeing constraints.
    \item[b] The total integration time represents only those frames that were included in the final combined mosaics.
    \item[c] The direct measurement of the FWHM from IRAF, independent of a profile model. 
      The derivative of the enclosed flux profile is computed and the peak of the azimuthally 
      averaged radial profile is found, where the FWHM is twice the radius of the profile at half the peak value. 
    \item[d] These are 5$\sigma$ depths in a circular aperture of 1.5$^{\prime\prime}$ diameter, which corresponds 
      to a total magnitude that is about 0.7 mag brighter for point sources for a seeing of 1.1$^{\prime\prime}$.
    \item[e] The aperture correction for a point-source to convert the 5$\sigma$ depths to total magnitudes.
    \end{tablenotes}
  \end{threeparttable}
  \label{tab:obs_table}
\end{table*}

Due to the dominance and short-term variability of the sky background, ground-based NIR imaging requires
many short dithered exposures.  This observing procedure is well established
\citep[see, e.g.,][]{Labbe03,Quadri07}, and facilitates the removal of artifacts remaining from bright
objects that can leave residual images in subsequent exposures.  The telescope follows a semi-random
dither pattern with a dither box of 90$^{\prime\prime}$, enabling good background subtraction without
significantly reducing the area with maximal exposure.  The dither box is sufficiently large to ``fill in'' the gaps 
between the four NEWFIRM arrays.

At each dither position, the exposure times were typically
1$\times$60 seconds (coadds $\times$ individual exposure time) for $J_{1}J_{2}J_{3}$, 3$\times$20 seconds for
$H_{1}H_{2}$ and 5$\times$12 seconds for $K$.  The internally-coadded frames make longer exposure times possible
at each dither position while staying in the linear regime of the array.

We obtained a total on-sky integration time between 17 and 31 hours in each filter in each field over the 3 semesters
(see Table~\ref{tab:obs_table} for details).  This represents all usable data, including aborted sequences or frames that
were outside weather specifications or seeing constraints; the total time on-sky was 294 hours.
When tallying only those frames that get included in
the final combined images, the total integration times range from 12 to 24 hours (a loss of $\sim$10--30\% of the data).

%=== Fig 2 
\begin{figure}[t!]
\leavevmode
\centering
\includegraphics[scale=0.45]{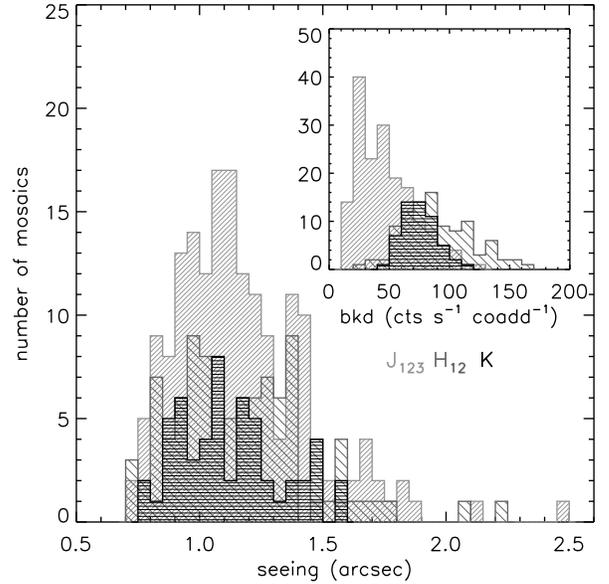}
\caption{The median seeing, as implied by a Moffat fit, and sky background in the combined sequences (typically $\sim$60 minutes) 
for $J_{1}, J_{2}$ and $J_{3}$ (light grey), $H_{1}$ and $H_{2}$ (grey), and $K$ (black).}
\label{fig:seeing}
\end{figure}

An overview of the median seeing and background levels measured in each observing sequence 
(ranging from 10--90 minutes, typically 60 minutes) is given in Figure~\ref{fig:seeing}.
The median seeing is measured from bright stars in the individual frames that are fit with 
a two-dimensional Moffat model using the robust, non-linear 
least squares curve fitting algorthm {\tt mpfit} in IDL~\citep{Markwardt09}.  
The data are of excellent quality with clean images that have no obvious systematic problems.

\section{Data Reduction}
\label{sec:reduction}

The data were reduced using a custom IDL code written by one of us (IL), following the method of the
IRAF\footnote{IRAF is distributed by the National Optical Astronomy Observatory, which is operated
by the Association of Universities for Research in Astronomy, Inc., under cooperative agreement with the
National Science Foundation.} package 
XDIMSUM\footnote{\burl{http://iraf.noao.edu/iraf/ftp/ftp/extern/xdimsum/}}.
The reduction process largely follows \citet{Labbe03},~\citet{ForsterSchreiber06},~\citet{Quadri07}
and~\citet{Taylor09b}.
Below, we summarize the process, highlighting modifications to the standard process described in the papers listed above.

\subsection{Image Processing}
\label{sec:processing}

Every individual raw frame was visually inspected to identify any poor quality frames.  To fall in this
catagory, the frame must either (1) contain artifacts such as satellite streaks, ghost pupils or other
transient features on the detector, (2) have a strongly varying background across the detector, (3) have
severe elongation of the point spread function (PSF) or (4) have double images of all objects 
due to ``jumps'' of the telescope or wind shake.  Between 10--30\% of the raw data was discarded for each filter,
resulting in final combined mosaics with negligible artifacts. 

The dark current is nonneglibible in NIR cameras and thus dark images with the appropriate exposure times
relative to the co-added science images are subtracted. Next, the raw images must be divided by the flat field
image to remove variations in the sensitivity across the array.

We constructed flat fields from dome flats in combination with observations of the open cluster M67,
accounting for both the small and large scale variations across the arrays, respectively.  
First, we determined the pixel-to-pixel response from dome flats.
The large scale structure in the dome flats was removed by median filtering using a 51$\times$51 pixel box, 
fitting a 7th order 2D Legendre polynomial to the median filtered image, and dividing by this fit. 
It was verified that subtracting rather than dividing by the fit changes the resulting residual image 
by $<$1\% (except for the $K$-band, where differences can be up to $\sim$5\% in the corners of the arrays).

The simple flat fields are not enough to correct to the large-scale sensitivity of the array due to 
imperfect illumination corrections.
We therefore determine the sensitivity variations across each array from observations of the 
open cluster M67 in an 8$\times$8 grid of positions spanning 28$^{\prime}\times$28$^{\prime}$; this way, 
many stars were observed on many array positions. The M67 data were dark subtracted, flat 
fielded using the ``flattened'' dome flat (i.e., the dome flat with large scale structure removed) and 
sky subtracted.  The fluxes of stars in M67 were measured using Source Extractor~\citep{Bertin96}, 
with a 4$^{\prime\prime}$ diameter aperture. Using the RA and DEC of the detected objects, sequences of unique stars 
imaged on multiple positions were identified.  From these stars, a response image was generated
reflecting the large scale response of the arrays. 
The scatter in the best-fit response was typically 1--3\%, with larger errors possible in small areas for
some array/filter combintations.
These ``response images'' were then multiplied by the “flattened domeflat” to create the flat fields. 

The individual dark-subtracted and flat-fielded frames are combined in two passes.  
During the first pass, the sky background is created for each science frame from a running median of the 
dithered sequence of the four preceeding and four subsequent science images.
Next, the dithered sky-subtracted images are cross-correlated with the central frame to determine the relative
shifts of the individual frames.  The relative flux scaling between the images is determined by measuring 
the flux of stars spread across the field.  Finally, the images are shifted to a common reference frame using
linear interpolation and are combined.  All objects in the combined image are detected using a simple thresholding
algorithm to generate an object mask.  This mask is then shifted back into the reference frame of each individual
science exposure and the background subtraction process described above is repeated in a second pass, using
a mean combine with cubic interpolation.  The second pass is optimal for the background subtraction as the objects are masked 
out during the calculation of the running median.  

%=== Fig 3
\begin{figure*}[t!]
\leavevmode
\centering
\includegraphics[scale=0.68]{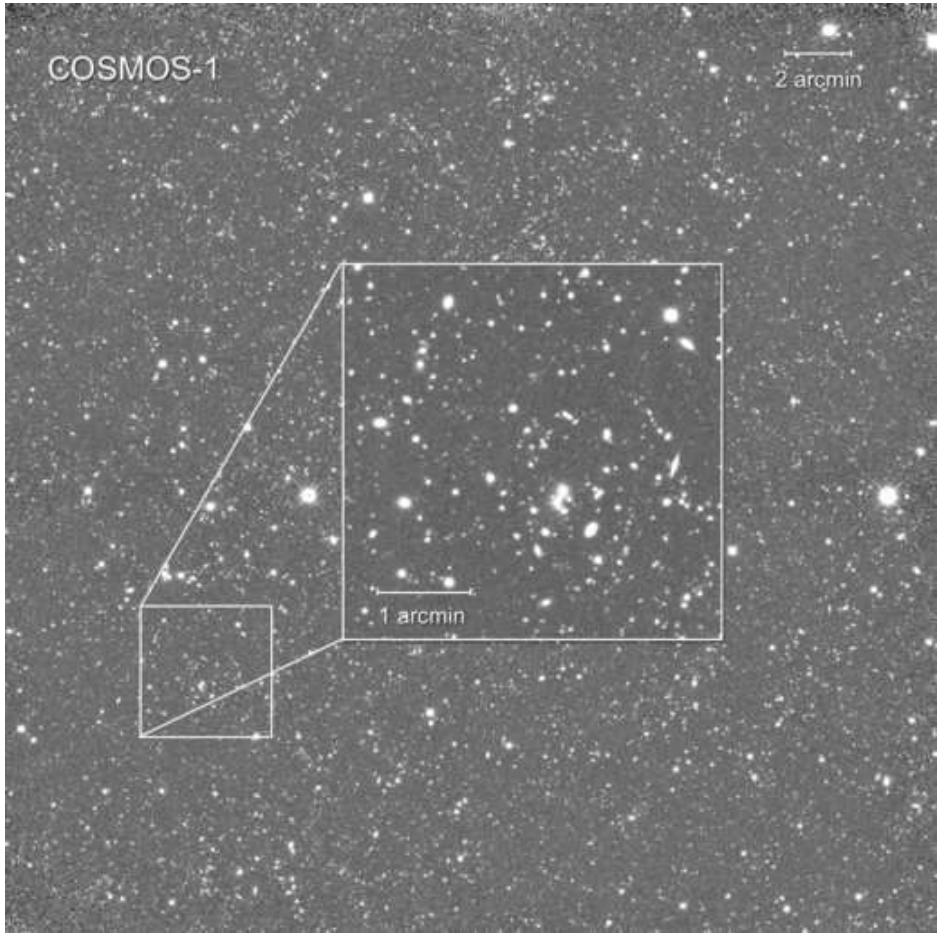}
\caption{The COSMOS $K$-band image, with a field of view of $\sim30\times30$ arcminutes.  The background is
both uniform and clean with no significant artifacts within the field that would effect the photometry.}
\label{fig:cosmosK}
\end{figure*}

%=== Fig 4
\begin{figure*}[t!]
\leavevmode
\centering
\includegraphics[scale=0.68]{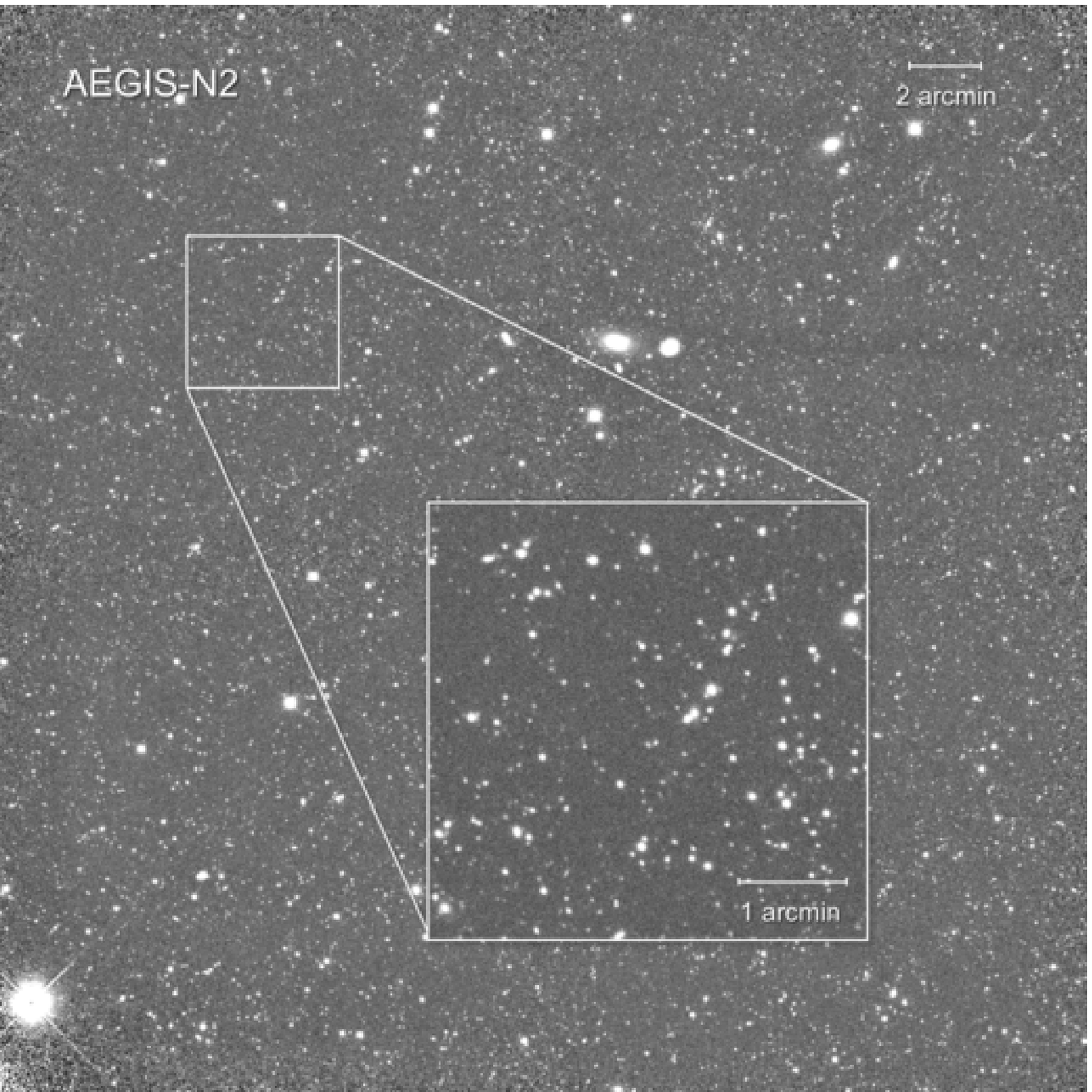}
\caption{The AEGIS $K$-band image (roughly $\sim30\times30$ arcminutes), with the inset demonstrating
the quality and depth of the data.}
\label{fig:aegisK}
\end{figure*}

\subsection{Further Optimization}
\label{sec:optimization}

In the reduction, we mask out any dead or 
hot pixels (or generically ``bad'' pixels) using a custom master bad pixel file.
Bad pixels have a non-linear response and these bad pixels can be easily separated from ``normal'' functioning
pixels during twilight.  The signal of good pixels increases
or decreases linearly with time, and so we fit the average flux of each quadrant as a function of the
time with a linear function.  Any dead or hot pixels will exhibit variable behavior,
resulting in drastically different best-fit slopes compared to properly function pixels. 
We flag any pixels with deviations $>$15\%.
The final bad pixel file contains a flag for any pixel that has been flagged in at least two of the twilight sequences,
in addition to the original bad pixel file from the NEWFIRM calibration library.  Although we
are conservative with the bad pixel file, the total fraction of bad pixels is still small ($\sim7$\%).  Any remaining 
bad pixels in each individual science frame are removed with a sigma-clipping algorithm before combining the images
within the reduction.

The nonlinearity of the NEWFIRM arrays increases with additional signal, 
reaching $\sim$5\% at 10,000 ADU.  Beyond
this level, the nonlinearity increases very steeply as the detector approaches saturation at 12,000 ADU.  
Coadding frames allows us to keep exposure times short thereby avoiding this nonlinear regime, 
as does observing when the sky levels are relatively low in a particular filter.  
A nonlinearity correction is applied to all individual frames, using the following empirical equation:

\begin{equation}
y=x\left(1 + \frac{6.64\times10^{-6}}{nx}\right)
\label{eq:nonlinear}
\end{equation}

\noindent where $x$ is the non-linear (raw) frame, $y$ is the non-linearity corrected frame, $n$ is the number of
internally coadded frames and $6.64\times10^{-6}$ is the average coefficient for the four 
arrays\footnote{\burl{http://www.noao.edu/staff/med/newfirm/fewompt/newfirm_linearity.pdf}}.  The non-linearity correction 
to the photometry is typically $<1$\% for the majority of frames.  

Another issue in NIR detectors is persistence images of bright objects, where the array has not completely
recovered from the previous exposures.  To alleviate this problem, we create a second object mask of the cores of the
brightest objects.  These are used to mask the pixels that contained the brightest objects in the previous two 
exposures in the final pass of the reduction.
The detectors also exhibit bias residuals which are constant along rows.  After masking all sources, 
the residual bias is removed by subtracting the median of each row in the individual sky-subtracted frames.  

The final major optimization of the image combination includes weighting the individual sky-subtracted frames
to maximize the signal-to-noise (S/N) ratio \citep[see e.g.,][]{Labbe03,Quadri07}.  
This method substantially improves the final image depth and quality
by assigning weights to the individual frames that take into account variations in the seeing, sky transparency,
background noise and PSF ellipticity.  To optimize the S/N,
the weight is proportional to the square of the flux scale (scale$_{i}$) divided by the 
median sky background (sky$_{i}$) and the median size of the seeing disk (FWHM$_{i}$), penalized for ellipticity:

\begin{equation}
w_{i}=\frac{\mathrm{scale}_{i}^{2}}{\mathrm{sky}_{i}\times\frac{\mathrm{FWHM}_{i}^{2}}{\sqrt{1-e^{2}}}}
\label{eq:weight}
\end{equation}

\subsection{Astrometry}
\label{sec:astrometry}

The weighted, sky-subtracted individual frames are combined into a single mosaic
with a pixel scale of 0.3$^{\prime\prime}$ pix$^{-1}$ (resampled from the native
NEWFIRM pixel scale of 0.4$^{\prime\prime}$ pix$^{-1}$).
The combined NMBS mosaics are aligned with the CFHT $I$-band images (see \S\ref{sec:ancillary}), with an astrometric precision of 
$\lesssim$0.1$^{\prime\prime}$ over the entire field of view.  The relative registration
between the NMBS and CFHT images must be precise to allow for accurate colors and cross-identification
of sources.  

The initial registration within the code is performed using a linear transformation in the first pass of the reduction.  
The rms variation in the position of individual sources is about 0.1--0.2$^{\prime\prime}$ (0.3--0.7 pixels) in
most cases, but can be as high as $\gtrsim$0.5--1$^{\prime\prime}$ (2--3 pixels) in
the corners of the arrays.  We fit the residual distortions with a second order polynomial, and combine this fit with the
original transformation. This combined distortion correction is then applied to the images, which are registered
using a cubic interpolation.  The resulting rms variation in 
position of the individual sources in the final combined images is $\lesssim$0.1$^{\prime\prime}$ (0.3 pixels).

\subsection{Photometric Calibration}
\label{sec:calib}

NIR spectro-photometric standard stars from the Calibration Database
System (CALSPEC)\footnote{\burl{http://www.stsci.edu/hst/observatory/cdbs/calspec.html}}
were observed in wide five-point dither patterns on photometric nights in the
2008A semester.  All of the other data were directly tied to the photometric data from 2008A.
Synthetic magnitudes of these stars were calculated by integrating
their observed (HST/NICMOS) spectra in the five medium-bandwidth filters~\citep[see Table 2 in][]{vanDokkum09a}. 
We adopt the mean zero points from these observations, listed in
Table~\ref{tab:obs_table}.  The zero points derived from these standard stars are remarkably
stable, with variations of $\lesssim$0.02 magnitudes throughout the observing program.  

A Galactic extinction correction has been applied to the photometry within the catalogs, as estimated from
\citet{Schlegel98}.  The corrections ranged from 4.5\% to 0.3\% for the $u$--$K$ photometry in
the AEGIS field, with slightly larger values of 8\% to 0.6\% in the COSMOS field due to its lower
Galactic latitude.

\subsection{Noise Properties}
\label{sec:noise}

The final combined $J_{1}$, $J_{2}$, $J_{3}$, $H_{1}$, $H_{2}$ and $K$ images constructed from the individually
registered, distortion-corrected, weighted averages of all frames are of excellent quality.  The flatness of the
background is readily visible in the final combined $K$-band images shown in Figures~\ref{fig:cosmosK} and~\ref{fig:aegisK}.  
Furthermore, the relatively deep and wide images are rich in compact sources.  
The combined images are slightly shallower near the edges and center of the image where there are gaps
between the NEWFIRM arrays, as these areas
receive less exposure time in the dither pattern.  This is reflected in the weight maps containing the
total exposure time per pixel.
The reduced NMBS images and weight maps will
be made publicly available through the NOAO archive\footnote{\url{http://archive.noao.edu/nsa/}}.  
 
Well-understood noise properties are important both for calculating the depths of the final combined images,
as well as measuring accurate photometric uncertainties.  Although the noise properties of the raw images are 
well described by the variance of the signal per pixel, the reduction process introduces correlations
between neighboring pixels.  Additionally, small errors in the background subtraction may contribute to the noise.

To estimate the noise, we follow~\citet{Labbe05} and empirically fit the dependence
of the normalized median absolute deviation (nmad) of the background as a function of the linear size $N$ of
an aperture with area $A$, where $N=\sqrt{A}$.  We measure the fluxes in $>$1000 independent circular apertures randomly 
placed on the registered, sky-subtracted, convolved images (see \S~\ref{sec:opticalpsf} for a description of the PSF matching) 
in a range of aperture sizes.  Any apertures that overlap with the detection image segmentation map are excluded.   

\begin{figure}[t!]
\leavevmode
\centering
\includegraphics[scale=0.47]{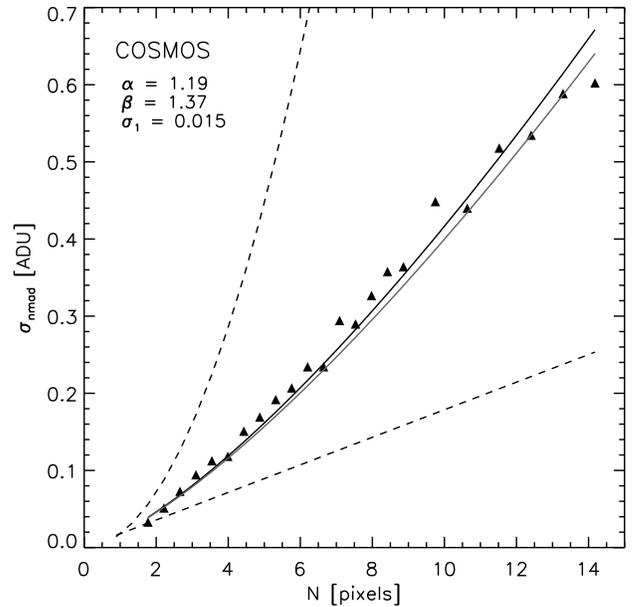}
\caption{The nmad value of background fluctuations for the COSMOS $K$-band image within an aperture with area $A$
as a function of the linear size $N$,
where $N=\sqrt{A}$.  The filled triangles illustrate the measured values, whereas the black solid curve is the
best fit to the data following the parametertization in Equation~\ref{eq:noise}.  The grey solid curve is the best fit
to the AEGIS data (not shown here).  The dashed curves show the expected scaling
relations of the background fluctuations in the case of no correlations amongst the pixels (bottom) and perfect
correlations of all pixels within the aperture (top).  Similar fits were performed for all filters in both fields.}
\label{fig:noise}
\end{figure}

As described in \citet{Quadri07}, the
scaling of the background noise with aperture size will be proportional to $N^{2}$ in the 
limiting case of perfect correlations between the pixels within an aperture.  In the other extreme case of uncorrelated adjacent pixels
within an aperture, the background noise scales with the linear size $N$.
The latter standard relation is unrealistic given the correlations between pixels that were introduced by the
imperfect background subtraction, but also by undetected sources, resampling of the pixels, etc.  The true scaling
between the background and the linear size falls somewhere between these two extremes.  Following 
\citet{Quadri07}, we parameterize the noise properties as:

\begin{equation}
\sigma_{\mathrm{nmad}} = \sigma_{1}\alpha N^{\beta},
\label{eq:noise}
\end{equation}

\noindent where $1<\beta<2$, $\sigma_{1}$ is the standard deviation of the background pixels and $\alpha$ is 
the normalization.  
In Figure~\ref{fig:noise}, we see that the width of the noise distribution within an 
aperture increases with increasing aperture size.  The best fit scaling of the background fluctuations
falls roughly between the two extremes of no correlation and a perfect correlation from pixel to pixel, 
with a best fit power-law index of $\beta=1.37$ in COSMOS ($\beta=1.35$ in AEGIS).  

%=== Fig 5

The background noise measurements are used to estimate the depths of the images.  In Table~\ref{tab:obs_table},
we list the 5$\sigma$ limiting depths.  These values are calculated using the background fluctuations within
a 1.5$^{\prime\prime}$ ``color'' aperture (see \S~\ref{sec:opticalpsf}).  
An additional aperture correction determined from the growth curves must be applied to these values to 
convert to total fluxes (see Column 7 of Table~\ref{tab:obs_table}).
The typical aperture correction for a point-source is 1.9 for a typical seeing of $\sim1.1^{\prime\prime}$.

\subsection{Ancillary Images}
\label{sec:ancillary}

The survey catalog combines the data taken with the NEWFIRM camera with publicly-available optical {\it ugriz}
broadband images of both survey fields from the Deep Canada-France-Haweaii Telescope Legacy Survey
(CFHTLS)\footnote{\burl{http://www.cfht.hawaii.edu/Science/CFHTLS/}}, using image versions
reduced by the CARS team \citep{Erben09, Hildebrandt09}, and $JHK_{S}$ broadband imaging
from the WIRCam Deep Survey (WIRDS; Bielby et al.,{\it in preparation}).
Additionally, we include deep {\it Subaru} images with the $B_{J}V_{J}r^{+}i^{+}z^{+}$ broadband 
filters and 12 medium-band filters in the COSMOS field \citep[][Y. Taniguchi et al. 2008, 
{\it in preparation}]{Taniguchi07}\footnote{\url{http://irsa.ipac.caltech.edu/data/COSMOS/images/}}.
We include mid-IR images in the {\it Spitzer}-IRAC bands over
the entire COSMOS pointing that are provided by the S-COSMOS survey \citep{Sanders07},
and partial coverage of the AEGIS pointing ($\sim0.15$ deg$^{2}$)
overlapping with the Extended Groth Strip \citep{Barmby08}.
The catalogs include {\it Spitzer}-MIPS fluxes at 24$\mu$m that are derived from images
provided by the S-COSMOS and FIDEL\footnote{\burl{http://irsa.ipac.caltech.edu/data/SPITZER/FIDEL/}}
surveys, with the same coverage as the IRAC data.
Finally, we incorporate NUV and FUV
{\it Galaxy Evolution Explorer}~\citep[GALEX;][]{Martin05} data in both fields from the Multimission Archive at Space Telescope
Science Institute (MAST) into the catalogs~\citep[see also][for a description of the COSMOS GALEX data]{Zamojski07}.

%=== Fig 6
\begin{figure*}[t!]
\leavevmode
\centering
\includegraphics[scale=0.55]{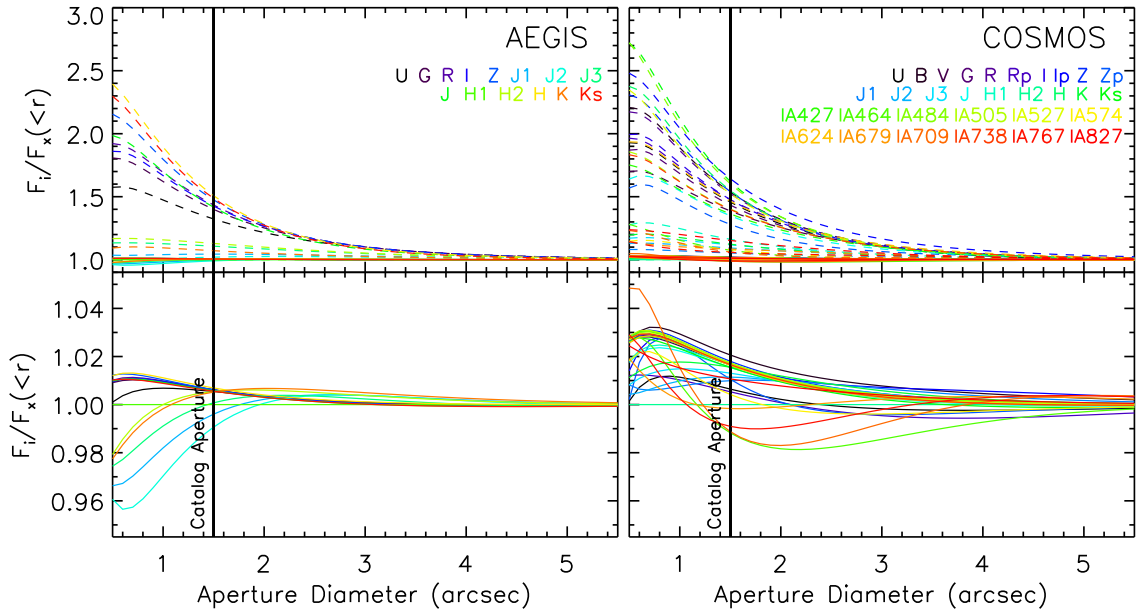}
\caption{The ratio of the growth curve for all optical to NIR bands $F_{\mathrm{i}}$ relative to the
band with the worst seeing $F_{\mathrm{x}}$ (where x = $H_{1}$ in AEGIS and $H_{2}$ in COSMOS).  The top
panel shows the ratio before convolution (dashed lines) and the bottom panel shows after convolution
(solid lines).  Note the different y-axis scales of the top and bottom panels.
The PSF matched images have accurate colors to $<2$\% for the catalog aperture diameter of
1.5$^{\prime\prime}$.}
\label{fig:growthcurve}
\end{figure*}

All ancillary images are matched to the same pixel scale (0.3$^{\prime\prime}$ pix$^{-1}$) and 
pointing of the NMBS images using the IRAF task {\tt wregister}.  The details of the source detection 
and extraction of photometry can be found in \S~\ref{sec:detection}.  In general, for objects that have 
flat spectra in $F_{\lambda}$, all of the ancillary data are deeper than the NMBS images with typical 
FWHM values of 0.8$^{\prime\prime}$ for the optical images.  For consistent photometry
in all (optical to NIR) bands, the images have all been convolved to the broadest PSF as determined
from the growth curves (see \S~\ref{sec:opticalpsf} for more details).
  
\section{Photometry and Source Detection} 
\label{sec:source}

We have 20 independent photometric filters in AEGIS, and in COSMOS we have 37 (FUV--8 $\mu$m).
The optical and NIR images were convolved to the same PSF before 
performing aperture photometry, so as to limit any bandpass-dependent effects.  
The source detection was done on the PSF matched images with Source Extractor \citep{Bertin96}
in relatively small apertures chosen to optimize the S/N (see \S~\ref{sec:noise}).  
To extract the photometry from the ultraviolet and mid-IR images,
we use an alternate source fitting algorithm especially suited for heavily confused images for which
a higher resolution prior (in this case, the $K$-band image) is available.

\subsection{Optical and NIR PSF Matching}
\label{sec:opticalpsf}

The FWHM of the PSF varies significantly between the optical and NIR bands: point sources in the 
optical images typically have FWHM of approximately 0.8$^{\prime\prime}$, whereas the point sources
in the medium-band NIR images have FWHM of 1.0$^{\prime\prime}$--1.2$^{\prime\prime}$.
Directly performing aperture photometry on these images would lead to different fractions of the flux 
falling within the aperture for different bands, thereby resulting in systematic color errors.

To minimize this problem, we degrade all of the optical and NIR images for each field to the 
broadest PSF.  We produce high S/N PSFs for each band from bright (unsaturated) stars covering the
entire field.  Square postage stamps (32 pixels on a side, or 9.6$^{\prime\prime}$) of the PSF 
stars were created, masking any nearby objects and averaged to create empirical PSFs.  
A kernel is then generated using the Lucy-Richardson method algorithm ({\tt lucy} in IRAF)
to convolve the empirical PSF to the broadest PSF ($H_{1}$ in AEGIS and $H_{2}$ in COSMOS).

Growth curves were determined for the empirical PSFs in each band, measuring the enclosed flux as a function of the aperture size.
In the top panels of Figure~\ref{fig:growthcurve}, we show the ratio of the growth curves of each band 
relative to the broadest PSF in both fields before degrading the PSF (dashed lines).  Generally,
the broadest PSFs (the slowest-growing growth curves) are the longer wavelength bands.  
The bottom panels of Figure~\ref{fig:growthcurve} show the ratio of the growth curves once they have been 
smoothed to the worst seeing.  For ``perfect'' PSF matching, this ratio should be unity at all aperture diameters.
At the aperture size used for the flux measurements (solid black line), the PSFs of all bands are 
matched within $<$2\% for both fields.

\begin{figure}[t!]
\leavevmode
\centering
\includegraphics[scale=0.47]{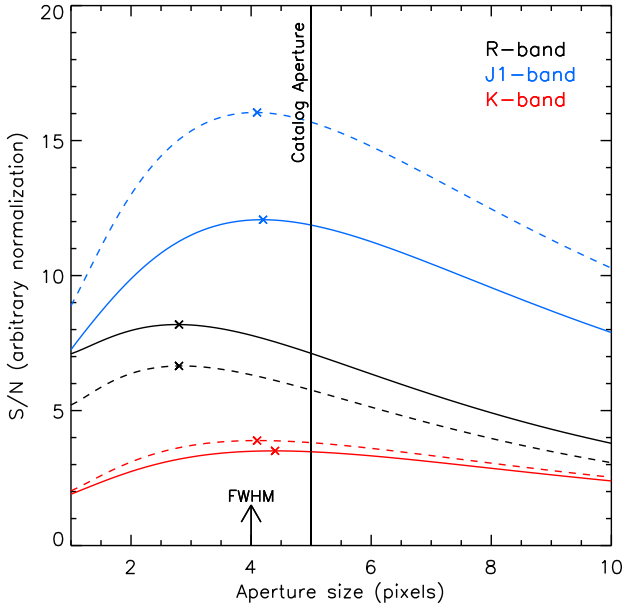}
\caption{The signal-to-noise as a function of the aperture size, derived from the ratio of the growth curve and
the noise properties for the $R$ (black), $J_{1}$ (blue) and $K$-band images (red).
The solid lines are for the COSMOS field and the dashed
lines are for AEGIS, with an ``x'' marking the peak in S/N.  The FWHM of the PSF matched $K$-band image
is shown with an arrow.  We use an aperture diameter of 5 pixels (1.5$^{\prime\prime}$) when
creating the NMBS catalogs so as to optimize the S/N while remaining larger than the FWHM.  }
\label{fig:SN}
\end{figure}

Given the noise properties of the images (see \S~\ref{sec:noise}) and the growth curve of the PSFs, 
we estimate the aperture size that optimizes the S/N for point sources in Figure~\ref{fig:SN}.
This optimization enables accurate color measurements which are necessary for photometric redshift calculations and
modeling of the stellar populations.
We adopt an aperture of 5 pixels (1.5$^{\prime\prime}$), defining the color aperture as:

\begin{equation}
F_{\mathrm{color},i}=\sum\limits_{\sqrt{x^2+y^2}<0.75^{\prime\prime}}F_{i}(x,y)
\label{eq:colap}
\end{equation}

\noindent where $F_{\mathrm{color},i}$ is the flux for each band $i$ 
within a circular aperture with a radius $r=\sqrt{x^2+y^2}$ of 0.75$^{\prime\prime}$.
The color aperture defined in Equation~\ref{eq:colap} is small enough to give nearly
optimal S/N for point sources and large enough to minimize possible systematics from residual astrometric or PSF matching
uncertainties.  
We note that the $K$-band has a relatively flat slope in S/N, whereas the
improvements are most prominent for the $J_1$ band (shown in Figure~\ref{fig:SN}).  There is a 20--30\% loss of
S/N when using larger apertures of 2--3${^{\prime\prime}}$ in bands blueward of the $K$-band.

%=== Fig 7                                                                                                      

The empirical estimate of the background noise $\sigma_{\mathrm{nmad}}$ described in \S~\ref{sec:noise} 
was repeated for all optical and NIR images within the color aperture of 1.5$^{\prime\prime}$.  
From these values, we estimate the flux uncertainies for each object in each band as:

\begin{equation}
\sigma_{\mathrm{color},i}^2=\left(\frac{\sigma_{\mathrm{nmad},i}}{\sqrt{\mathrm{w_i}/\mathrm{w}_{\mathrm{median},i}}}\right)^2+\frac{F_{\mathrm{color},i}}{g_i}
\label{eq:error}
\end{equation}

\noindent The photometric error for each band $i$ is the sum in quadrature of the background uncertainty ($\sigma_{\mathrm{nmad},i}$; 
defined in Equation~\ref{eq:noise}) weighted by the fractional exposure $\mathrm{w_i}$ at each object's location 
relative to the median $\mathrm{w}_{\mathrm{median},i}$,
and the source Poisson noise $F_i/g_i$ for each band, with the former dominating the noise for faint
sources and the latter for bright sources.  $F_i$ is the 
flux of the object in ADU within a 1.5$^{\prime\prime}$ aperture
and $g_i$ is the total effective gain.  Note that these are the uncertainties of the aperture fluxes, 
not the total fluxes.  The uncertainty in the total fluxes listed in the catalogs is:

\begin{equation}
\sigma_{\mathrm{total},i}=\sigma_{\mathrm{color},i}\times\frac{K_{\mathrm{total}}}{K_{\mathrm{color}}}
\label{eq:toterror}
\end{equation}

The $K$-band total magnitude is calculated from the SExtractor AUTO photometry (see Equation~\ref{eq:Ktot}),
with an uncertainty calculated as follows:

\begin{equation}
\sigma_{\mathrm{Ktot}}^{2}=\left(\frac{\sigma_{1}\alpha\left(\pi R_{\mathrm{Kron}}^{2}\right)^\frac{\beta}{2}}{\sqrt{\mathrm{w}_K/\mathrm{w}_{\mathrm{median},K}}}\right)^{2}+\frac{K_{\mathrm{AUTO}}}{g_K}
\label{eq:Ktoterror}
\end{equation}

\noindent where $R_{\mathrm{Kron}}$ is the circularized Kron radius~\citep{Kron} and $\sigma_{1}$, $\alpha$ and $\beta$
are defined in Equation~\ref{eq:noise}, weighted by the fractional exposure time.  
The circularized Kron radius is calculated from the semimajor and semiminor
axes as $R_{\mathrm{Kron}}=\sqrt{\mathrm{A\_IMAGE}\times\mathrm{B\_IMAGE}\times\mathrm{KRON\_RADIUS}^2}$, 
using the output SExtractor parameters.  The following equation can be used to determine 
the full error in the total magnitude for a single band in isolation, including systematics:

\begin{equation}
\sigma_{\mathrm{full},i}^2=\sigma_{\mathrm{total},i}^2+\sigma_{\mathrm{Ktot}}^2-\sigma_{\mathrm{total},K}^2
\label{eq:fullerr}
\end{equation}

The derivation of the uncertainties
for the ultraviolet and mid-IR data are described in \S~\ref{sec:iracpsf} and \S~\ref{sec:uvpsf}.

\subsection{Source Detection}
\label{sec:detection}

The sources are detected in a noise-equalized $K$-band image
using SExtractor in dual-image mode, where the aperture photometry 
is simultaneously measured on the PSF matched $u$--$K$ images.  In both fields, 
the $K$-band detection image is the unsmoothed (pre-PSF matching) NEWFIRM $K$-band image multiplied by the 
square root of the weight map (based on fractional exposure time per pixel).
The input SExtractor parameters were optimized so as to find all faint objects while limiting the number of
spurious detections.  The optimal setting for the minimum number of pixels above the threshold to trigger a detection 
was 5 pixels, with a detection threshold of 1.2$\sigma$ and 32 deblending sub-thresholds.
We use a Gaussian filter for the detections with a FWHM of 4 pixels, first performing a global background subraction
in all optical--NIR images with a background mesh size of 32 pixels and a smoothing filter of 3 pixels. 
We note that we used a minimum Kron radius of 3 pixels, compared to the default value of 3.5 pixels. 
A fixed aperture diameter of 1.5$^{\prime\prime}$ (5 pixels) was used to both enclose a significant
fraction of a given object's flux while minimizing uncertainties due to background fluctuations 
(see \S~\ref{sec:opticalpsf}, Figure~\ref{fig:SN}).  

Total fluxes were determined using the SExtractor AUTO photometry, which uses a flexible elliptical aperture.
This aperture contains the majority of the flux for bright objects but may be missing a large fraction of the light 
for fainter objects.  Therefore an aperture correction is applied to convert the AUTO flux to total flux 
\citep[see, e.g.,][]{Labbe03}.  The $K$-band total flux is calculated as follows,

\begin{equation}
K_{\mathrm{total}}=K_{\mathrm{AUTO}}\times\frac{1}{f_{\mathrm{R<AUTO}}}
\label{eq:Ktot}
\end{equation}

\noindent where $K_{\mathrm{AUTO}}$ is the AUTO flux and $f$ is the fraction of light enclosed within the 
$K$-band growth curve generated during the PSF matching described 
in \S~\ref{sec:opticalpsf} for a circular aperture with the same area as the AUTO aperture.  
We force a minimum size of the AUTO aperture equal the size of the color aperture, affecting $<0.1$\% of the sources.  
The aperture correction to the AUTO fluxes is typically on the order of 5--10\%, extending to higher values for 
the faint, extended sources.  The total flux for each band $i$ is calculated as follows:

\begin{equation}
F_{\mathrm{total},i}=F_{\mathrm{color},i}\times\frac{K_{\mathrm{total}}}{K_{\mathrm{color}}}
\end{equation}

\noindent where $F_{\mathrm{color},i}$ is the flux within a color aperture of 1.5$^{\prime\prime}$ diameter (see Equation~\ref{eq:colap}).

\subsection{IRAC/MIPS PSF Matching and Photometry}
\label{sec:iracpsf}

The FWHM of the IRAC and MIPS images are a factor of $\sim$2 and $\sim$5 broader than the PSF-matched images.
Furthermore, the IRAC and MIPS images contain significant non-Gaussian
structure that results from point source diffraction.  Smoothing the optical to NIR photometry
to the IRAC PSF shape would substantially reduce the S/N and cause significant blending issues in 
the deep optical data in addition to the IRAC/MIPS images.  

We use a source-fitting algorithm developed by one of us (IL) that is designed for heavily confused images to extract the
photometry from the IRAC and MIPS images \citep[see, e.g.,][]{Labbe06, Wuyts07, Marchesini09, Williams09}.
As sources that are bright in $K$ are also
typically bright in the IRAC images (and to a lesser extent MIPS images), the $K$-band can be used
as a high resolution template to deblend the IRAC/MIPS photometry.       
The information on position and extent of the sources 
based on the higher resolution $K$-band segmentation map is therefore used to model the lower resolution 
IRAC 3.6--8.0 $\mu$m and MIPS 24 $\mu$m images.  This technique is similar to \citet{Laidler07}, with one important
difference.  \citet{Laidler07} use the best-fit fluxes directly to calculate flux ratios (colors), whereas
our method adds back in the residuals of the fit before performing aperture photometry.

Each source was extracted separately from the $K$-band image and, 
under the assumption of negligible morphological K-corrections, convolved to the IRAC/MIPS resolution 
using local kernel coefficients.  The convolution kernels were constructed using bright, isolated, 
unsaturated sources in the $K$ and IRAC/MIPS bands, derived by fitting a series of Gaussian-weighted 
Hermite functions to the Fourier transform of the sources.  Outlying or poorly-fit kernels are 
rejected and a smoothed 2D map of the kernel coefficients was stored.  All sources in each IRAC/MIPS 
image are fit simultaneously, with the flux left as a free parameter.  The modeled light of 
neighboring objects is subtracted, thereby leaving a clean IRAC/MIPS image to perform aperture 
photometry in fixed 3$^{\prime\prime}$ apertures for IRAC and 6$^{\prime\prime}$ apertures for MIPS.
For an illustration of this technique, see Figure 1 in \citet{Wuyts07}.  

%=== Fig 8                                                                                                        
\begin{figure*}[t!]
\leavevmode
\centering
\includegraphics[scale=0.7]{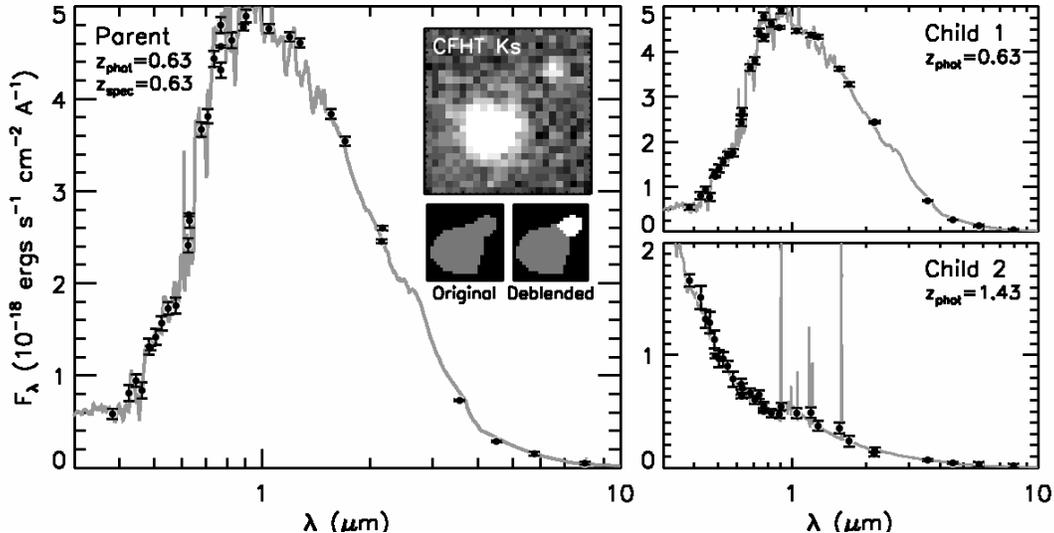}
\caption{An example of the SED of a galaxy that is blended in the original catalog (left) compared to
the SEDs of the deblended children (right).  The inset panel shows the CFHT $K_{S}$-band image with the original
and deblended segmentation maps below. The SEDs of the 2 children are very different.  The photometric redshift
of the first child is usually very similar to that of the parent.}
\label{fig:deblendsed}
\end{figure*}

In order to compute a consistent $K$-IRAC color,
we measure the flux of the objects on a cleaned $K$-band image convolved to the IRAC resolution within the 
same aperture to correct the photometry for flux that falls outside of the aperture due to the 
large PSF size.  We then scale the photometry to total fluxes 
by measuring the ratio of the total $K$-band flux to the
convolved $K$-band flux within a 3$^{\prime\prime}$ aperture for each object:

\begin{equation}
F_{\mathrm{IRAC,catalog}}=F_{\mathrm{IRAC,cleaned}}(3^{\prime\prime})\times\frac{K_{\mathrm{total}}}{K_{\mathrm{conv}}(3^{\prime\prime})}
\label{eq:irac}
\end{equation}

\noindent where $F_{\mathrm{IRAC,cleaned}}$ is the flux of the object measured with the modeled light of the 
neighbors removed, $K_{\mathrm{conv}}$ is the flux of the object measured in the $K$-band image after smoothing
it to the IRAC resolution, and $K_{\mathrm{total}}$ is the total $K$-band flux.
The $F_{\mathrm{IRAC}}$ catalog flux is the total flux that would have been measured in a 1.5$^{\prime\prime}$ 
aperture with an aperture correction if IRAC had the same PSF as the other data.

The uncertainties in the measured fluxes are derived from the residual contamination of the subtracted neighbors.  
The normalized convolved $K$-band image of each neighboring source is scaled by the formal 1$\sigma$ error in the 
fitted flux as computed from the covariance matrix produced by least-squares minimization.  By performing aperture 
photometry on the image of the residual neighbor contamination, we get the error corresponding to 
the source flux measured within an aperture of the same size.

The catalog includes the 24 $\mu$m flux within fixed 6$^{\prime\prime}$ apertures and total fluxes.
To measure the total 24 $\mu$m fluxes, we create a MIPS growth curve from several bright, isolated, unsaturated point sources
within each field.  These square postage stamps are 12.6$^{\prime\prime}$ on a side, and we derive an aperture correction
from 6$^{\prime\prime}$ to 12.6$^{\prime\prime}$ of a factor of 2.8.  To convert to total flux, we include an additional aperture
correction for the 17\% of the flux that falls outside 
13$^{\prime\prime}$\footnote{\burl{http://ssc.spitzer.caltech.edu/mips/mipsinstrumenthandbook/}}.

\subsection{GALEX PSF Matching and Photometry}
\label{sec:uvpsf}

The GALEX Far UV band (1350--1750\AA) and Near UV band (1750--2800\AA) have 4.5$^{\prime\prime}$--6$^{\prime\prime}$
resolution (FWHM), roughly $\sim$4--5 times broader than the PSF matched optical to NIR photometry.
The same method described in \S~\ref{sec:iracpsf} is used to extract the GALEX photometry.
In this case, however, the higher resolution prior is the CFHT $u$-band image.  

As the $u$-band contains many blue galaxies not in the $K$-band detection image, 
a new segmentation map was generated with SExtractor for the $u$-band and matched to
the $K$-selected catalog.  This allows for all of the objects within the field of view to be modeled to clean the GALEX
images for aperture photometry, regardless of whether these objects are in the NMBS $K$-selected catalog.

Following \S~\ref{sec:iracpsf}, we measure an aperture correction from the flux of the objects on 
the cleaned $u$-band image convolved to the GALEX resolution in order to compute a consistent GALEX-$u$ color.  
We then scale the photometry to the same color apertures used for the
optical to NIR photometry by measuring the ratio of the convolved $u$-band flux within a 6$^{\prime\prime}$
aperture and our catalog aperture size for each object, analogous to Equation~\ref{eq:irac}.

\subsection{Deblending}
\label{sec:deblend}

A significant fraction of the objects detected by SExtractor are
actually blended objects, although classified as a single object. 
Although SExtractor implements a deblending algorithm, there is no set
of parameters that will enable the splitting of close pairs while also detecting
faint objects, which requires pre-smoothing.  Deblending the photometry
is particularly important when studying the most luminous galaxies; the fraction
of sources that are deblended increases from $\sim10$\% at $K$=22 mag to $\sim40$\%
at $K$=20 mag, reaching as high as $\sim80$\% of sources with $K=18$ mag.
In this section, we describe the method used to deblend the photometry
of the single ``parent'' object into its constituent ``children'',
based on the technique described by~\citet{Lupton01}.

The deblending is done using the WIRDS $K_S$-band image (0.7$^{\prime\prime}$ seeing) in COSMOS
and a combination of the WIRDS $K_S$-band image (0.7$^{\prime\prime}$ seeing) where the exposure is 
greater than 30\% of the maximum and the non PSF matched $K$-band image in AEGIS (1.1$^{\prime\prime}$)
in the remaining area.  Using this criterion, roughly half of the AEGIS deblending image is from the 
higher resolution WIRDS data.  
Peaks are found after applying a Difference of
Gaussians band pass filter that preserves scales similar to the 
seeing size. Object templates are created by comparing pairs of 
pixels symmetrically around the peak and replacing both by the lower value.
False peaks are rejected if the total template
S/N$<3$ or if the template peak value is less than 2$\times$
the combined pixel value of all other templates at that location.
Finally, we solve for the template weights by a linear fit of the 
templates to the image, and then assign flux to every pixel of each 
child according to the relative contribution of the weighted templates.
The segmentation map is updated, assigning each pixel to the
child with the largest contribution.

Using the peaks previously found in the high resolution image, the 
deblending procedure is repeated for each $u$–-$K$ background subtracted, 
PSF matched image. Aperture photometry in 1.5$^{\prime\prime}$ color apertures is then 
performed on the ``deblended'' children. Total fluxes are computed by 
dividing the parent total flux according to the aperture flux ratio of the
children, thus conserving the total flux of the parent. 
The IRAC and MIPS photometry of the children is remeasured using the method described 
in \S\ref{sec:iracpsf}, but now using the new deblended segmentation map and the 
positions of the children. Due to the large PSF size of the GALEX data and 
the lack of a deblended $u$-band segmentation map, no attempt was made
to deblend the NUV and FUV data.

%=== Fig 9                                                                                                       
\begin{figure*}[t!]
\leavevmode
\centering
\includegraphics[scale=0.7]{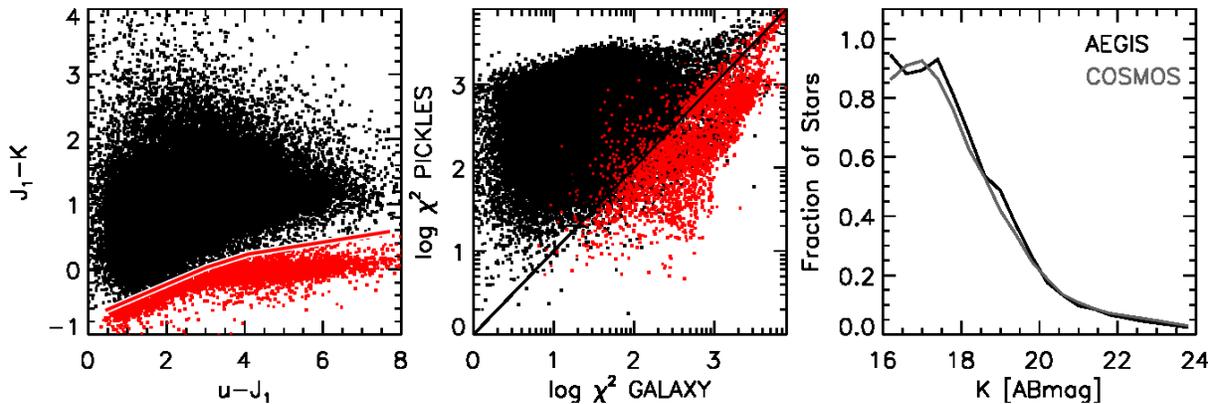}
\caption{Objects are classified as stars (red) from their $u$--$J_{1}$ and $J_{1}$--$K$ colors in the
left panel.  These same objects also have SEDs that tend to be better-fit by the stellar templates
than the galaxy templates (middle panel).  All objects in both catalogs are shown, with no
quality cuts applied.  The fraction of stars relative to the total number of
objects as a function of $K$-band magnitude for both fields is shown in the right panel.}
\label{fig:starflag}
\end{figure*}

2186 and 3017 objects in the original catalogs were split into 5019 and 6803 
children in AEGIS and COSMOS, respectively. New identification 
numbers of the children start at ID=30000.  On average, the difference between 
the total $K$-band flux of the brightest child relative to the parent 
increases with redshift, with a median difference of 0.1 mag at z $\sim 0.5$ 
and 0.3-–0.4 mag at $z>1.5$.  

Figure~\ref{fig:deblendsed} shows an example of an object that has been deblended with the above 
algorithm. Although classified as a single object by SExtractor, there is clearly a
faint neighbor. The deblending technique recovers a well-described SED of the 
faint neighbor, with a notably different shape and photometric redshift compared 
to the brighter galaxy. It is worthwhile to note that, in case of a fainter neighboring 
object (e.g., child 2 in Figure~\ref{fig:deblendsed}), the photometric redshift of the brighter object (child 1) 
is generally not significantly different from the parent, even though the children 
can have vastly different spectral shapes and photometric redshifts.

\subsection{Star Classification}
\label{sec:starflag}

It is important to distinguish between galaxies and objects that are likely stars in the photometric 
catalog.  Stellar SEDs have a characteristically steep rise and fall in their spectra, resulting
in a tight stellar sequence in their $u$--$J$ and $J$--$K$ colors relative to the colors of galaxies.
This is shown in Figure~\ref{fig:starflag}: stars (red) form a distinctive sequence with blue $J$--$K$ colors for 
a wide range of $u$--$J$ colors as compared to galaxies (black).

Stars are flagged based on the $u$--$J_{1}$ and $J_{1}$--$K$ colors of all objects, where any objects
satisfying the following limits (corresponding to the red line in Figure~\ref{fig:starflag}) will be classified as a star:

\begin{eqnarray}
J_{1}-K < -0.74+0.26(u-J_{1})~~~~~~~~~~[&u-J_{1}&<3]\nonumber\\
J_{1}-K < -0.55+0.19(u-J_{1})~~~~~[3<&u-J_{1}&<4]\\
J_{1}-K < -0.16+0.10(u-J_{1})~~~~~~~~~~[&u-J_{1}&>4]\nonumber
\label{eq:starflag}
\end{eqnarray}

The middle panel in Figure~\ref{fig:starflag} shows the $\chi^{2}$ values for the stars (red) and galaxies (black) 
when fit with \citet{Pickles98} stellar template library compared to the default set of EAZY galaxy 
templates.  The vast majority of objects that are classified as stars through their $u$--$J$ and $J_{1}$--$K$
colors are better fit with the stellar templates.
The fraction of objects that are classified as stars relative to the total number of objects as a function
of $K$-band magnitude is shown in the right panel in Figure~\ref{fig:starflag} for both fields.
Almost all objects brighter than $K\sim16$ mag are stars.  

We have compared our star classification to the COSMOS {\it Hubble Space Telescope} (HST) Advanced Camera for Surveys (ACS)
catalog\footnote{\url{http://irsa.ipac.caltech.edu/data/COSMOS/datasets.html}}, matching objects within a radius of 1$^{\prime\prime}$.  
We find that $>90$\% of the objects classified as stars using the higher resolution $I$-band data also fall within 
our $u$--$J_{1}$ and $J_{1}$--$K$ color selection limit in Equation~\ref{eq:starflag}.
The star classification method seems robust, although we note that there may be a small fraction of 
objects ($<10$\% of stars and $<1$\% of galaxies) 
that are misclassified, especially close to the detection limit. 

\subsection{Catalog Format}
\label{sec:catalog}

The final $K$-selected catalog consists of 24739 and 27520 objects detected
by SExtractor in the AEGIS and COSMOS mosaics, with a total of 27572 and 31306 objects in the deblended catalogs.  
Photometry in the FUV through the MIPS 24 $\mu$m
band is included, with additional optical broad- and medium-band photometry in the COSMOS field.
The catalog fluxes can be converted into total magnitudes as follows:  

\begin{equation}
m = -2.5\log{F}+25
\label{eq:flux}
\end{equation}

\noindent where $F$ is the total flux density of the object with a magnitude $m$ normalized to a 
zero point of 25 in the AB system, corresponding to a flux density of $3.631\times10^{-30}$ erg s$^{-1}$ Hz$^{-1}$ cm$^{-2}$
(or 0.3631 $\mu$Jy).  
To convert from total magnitudes to the color aperture magnitude,
the fluxes and errors should be multiplied by the ratio of the $K$-band aperture flux ({\tt Kaper}) to the
total $K$-band flux.  
A description of the columns included in the Version 5.0 catalogs can be found in Table~\ref{tab:catalog}.

A standard selection of galaxies can be obtained by selecting {\tt use}=1 for either
the original SExtractor catalog or the deblended catalog, 
which is equivalent to {\tt wmin}$>$0.3 and {\tt star\_flag}$=$0. 
An additional cut on the S/N ratio of the galaxies may be optimal.

\begin{table*}[t!]
  \caption{Summary of Photometric Catalog Contents}
  \centering
  \begin{threeparttable}
    \begin{tabular}{lll}
      \hline
      \hline
      Column No. &  Column Title  & Description \\
      \hline
      1   & id & Object Indentifier, beginning from 1 \\
      2,3 & x, y & x- and y-image coordinates \\
      4,5 & ra, dec & Right ascension and declination (J2000.0; decimal degrees) \\
      6,7 & Kaper, eKaper & Color aperture $K$-band flux and error \\
      8--82 (8--67)\tnote{a} & X, eX, wX & Total flux, error, and weight in each filter  \\
      83--106 & X, eX & Total flux and error of the COSMOS optical medium-bands \\
      107 (68) & ftot24um\_uJy & Total MIPS 24 $\mu$m flux ($\mu$Jy) \\
      108, 109 (69, 70) & f24um\_uJy, e24um\_uJy & Aperture flux within 6$^{\prime\prime}$ and error of MIPS 24 $\mu$m ($\mu$Jy) \\
      110 (71) & w24um & Weight of MIPS 24 $\mu$m \\
      111 (72) & wmin & Weight of $u$--$K$ band with minimum exposure \\
      112 (73) & wmin\_irac & Weight of IRAC band with minimum exposure \\
      113 (74) & z\_spec & Spectroscopic redshift, if available (matched within 1$^{\prime\prime}$ radius) \\
      114 (75) & star\_flag & Star flag (1=Star, 0=Galaxy) \\
      115, 116 (76, 77) & ap\_col, ap\_tot  & Size of the color aperture and total aperture in arcseconds \\
      117 (78) & totcor & Aperture correction applied to the $K$-band AUTO flux to convert to total flux \\ 
      118, 119 (79, 80) & K\_ellip, K\_theta\_J2000 & $K$-band ellipticity and position angle from SExtractor \\
      120 (81) & K\_R50 & $K$-band half-light radius from SExtractor (pixels) \\
      121 (82) & K\_class\_star & $K$-band SExtractor parameter measuring stellarity of object \\
      122, 123 (83, 84) & K\_flags, UH2\_flags & $K$-band detection flag and maximum detection flag from $u$--$H_{2}$ from SExtractor\\
      124 (85) & Near\_Star & Signifies object is near a bright star and photometry may be affected \\
      125--130 (86--91) & X\_contam & Ratio of the IRAC/GALEX flux in neighboring objects removed to the ``cleaned'' flux \\
      131 (92) & contam\_flag & IRAC/GALEX contamination ratio $>$50\% in any band (1=$>$50\% in $\geq$1 band, 0=OK)\\
      132, 133 (93, 94) & nchild, id\_parent & Number of children from deblended catalog (-1 if child) and ID of parent (-1 if parent) \\
      134 (95) & use & Standard selection of galaxies for either the deblended or the original catalogs (use = 1)  \\
      \hline
    \end{tabular}
    \begin{tablenotes}
    \item[a] AEGIS column numbers are in parentheses, as the COSMOS catalog contains additional bands
    \end{tablenotes}
  \end{threeparttable}
\label{tab:catalog}
\vspace{0.7cm}
\end{table*}

\subsection{Completeness}
\label{sec:completeness}

%=== Fig 10                                                                                          
\begin{figure}[t!]
\leavevmode
\centering
\includegraphics[scale=0.48]{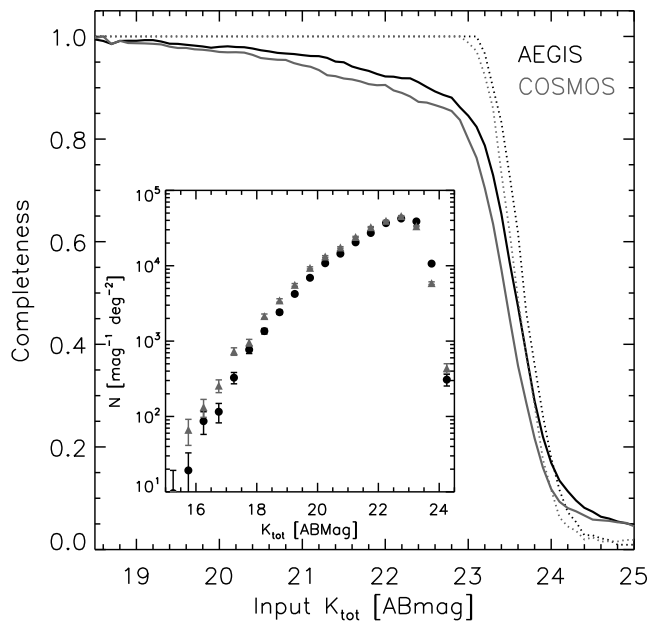}
\caption{$K$-band completeness curves; the completeness is defined as the fraction of 
simulated point sources that were recovered as a function of the total $K$-band magnitude
of the simulated source.  The point sources were inserted at random locations across the entire
image where the exposure map was greater than half the maximum exposure, with no masking (solid lines)
and detected sources masked (dotted lines).  
The inset plot shows the raw number counts of galaxies within the two NMBS fields shown with  
Poisson error bars, with no corrections for incompleteness.  }
\label{fig:completeness}
\end{figure}

\begin{table}[b!]
  \caption{Point-Source Completeness Limits}
  \centering
  \begin{threeparttable}
    \begin{tabular}{lcccc}
      \hline
      \hline
      & \multicolumn{2}{c}{Masking Sources} & \multicolumn{2}{c}{Entire Image} \\
      \cline{2-5}
      Field & 90\% Limit\tnote{a} & 50\% Limit & 90\% Limit & 50\% Limit \\
      \hline
      AEGIS  & 23.2 & 23.7 & 22.5 & 23.6 \\
      COSMOS & 23.2 & 23.6 & 22.1 & 23.4 \\
      \hline
    \end{tabular}
    \begin{tablenotes}
    \item[a] All magnitudes are given in the AB system.
    \end{tablenotes}
  \end{threeparttable}
%    \footnotetext[1]{Magnitudes are given in the AB system.}
  \label{tab:completeness}
\end{table}

We estimate the (point-source) completeness of our catalogs as a function of magnitude by attempting to 
recover simulated sources within the $K$-band noise-equalized detection images.  The
simulated point sources were generated from the $K$-band empirical PSF described in 
\S~\ref{sec:opticalpsf}, before smoothing.  The empirical PSF is scaled to the desired
flux level and inserted at random locations within the entire image.  We then run
SExtractor using the same settings described in \S~\ref{sec:detection}, determining
the fraction of recovered sources as a function of input $K$-band total magnitude.  
Figure~\ref{fig:completeness} shows the resulting completeness curves from this analysis.
The completeness curves qualitatively agree with the raw number counts of galaxies within the fields
shown in the inset plot. 

A fraction of the simulated point sources will fall on or close to other objects and these
blended objects may not be properly handled by SExtractor.  We have also repeated the simulations
by inserting the point sources in locations that do not overlap with the segmentation map, thereby
``masking'' the original detected sources (dotted lines in Figure~\ref{fig:completeness}).  
We include the 90\% and 50\% completeness limits for
both the masked and unmasked simulations in Table~\ref{tab:completeness}.  We note that about 10\% of the objects
are blended within the original catalogs produced by SExtractor, as determined by the additional
deblending algorithm described in \S~\ref{sec:deblend}.  When using the deblended catalogs, these
completeness limits may therefore be slightly conservative.  As the NMBS $K$-band
fields are not overly crowded, the completeness limits for both simulations are similar.

\subsection{Number Counts}
\label{sec:counts}

%=== Fig 11   
\begin{figure}[t!]
\leavevmode
\centering
\includegraphics[scale=0.48]{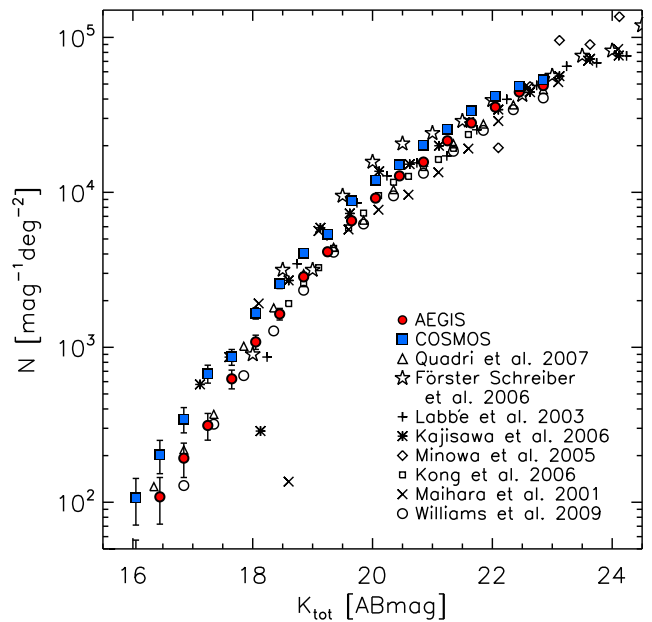}
\caption{Average galaxy number counts from the two NMBS fields, along with a compilation of results
from the literature.  Poisson error bars are shown for the NMBS fields; the other fields have comparable
uncertainties.  The NMBS fields have been corrected for incompleteness, but only points where this 
correction is $\lesssim20$\% are shown.}
\label{fig:counts}
\end{figure}

Figure~\ref{fig:counts} shows the surface density of galaxies as a function of magnitude in the NMBS
fields compared to other galaxy number count analyses drawn from the literature.  Objects
that are classified as stars (see \S~\ref{sec:starflag}) have been excluded from this calculation.
The number counts are calculated in 0.4 mag bins using objects within the image area with $>$30\% 
of the total $K$-band exposure time ($\sim$0.21 deg$^{2}$ in both fields).  

The number counts of the
NMBS fields have been corrected for incompleteness using the simulations without masking, as calculated in \S~\ref{sec:completeness}.
We note that the completeness values only provide a simplistic correction to the surface density of 
objects, whereas a more sophisticated analysis would account for the difference between measured
and intrinsic fluxes of the synthetic sources.  Due to the potential biases in the photometry of the faintest
sources, we only extend the average number counts to $K\sim22.7$, where the completeness correction 
begins to become significant (Figure~\ref{fig:completeness}).

The number counts of the NMBS fields agree with the large compilation of data sets, within the scatter
due to field to field variations.  We note that the COSMOS field appears to have a relatively high density of 
objects at $K<18$ mag.  We refer the reader to \citet{Wake11} for a detailed 
analysis of the relative densities and galaxy clustering in the NMBS fields.

\subsection{Comparison to Other Photometric Catalogs}
\label{sec:othercatalogs}

%=== Fig 12
\begin{figure}[b!]
\leavevmode
\centering
\includegraphics[scale=0.46]{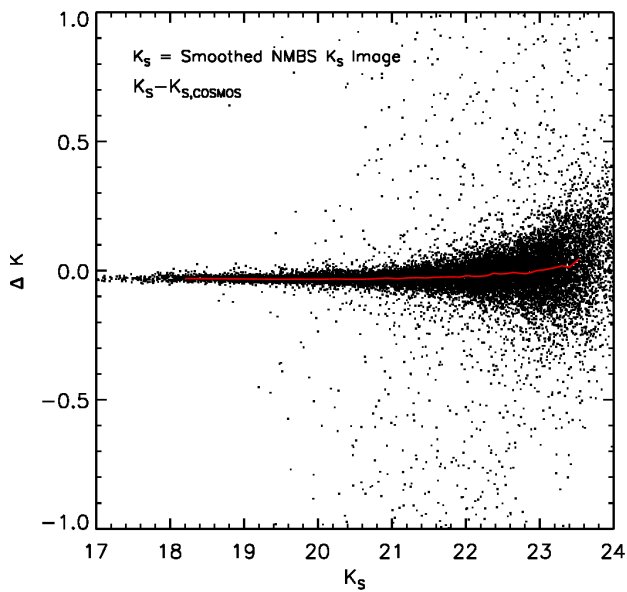}
\caption{The difference in $K$-band magnitude between the NMBS $K_{S}$ image smoothed with a Gaussian to 
1.5$^{\prime\prime}$ seeing and the COSMOS catalog magnitude within 3$^{\prime\prime}$ apertures (black points).
The running median (red) indicates that there are no significant systematic zero point offsets between the two catalogs.}
\label{fig:compareK}
\end{figure}

%=== Fig 13
\begin{figure*}[t!]
\leavevmode
\centering
\includegraphics[scale=0.72]{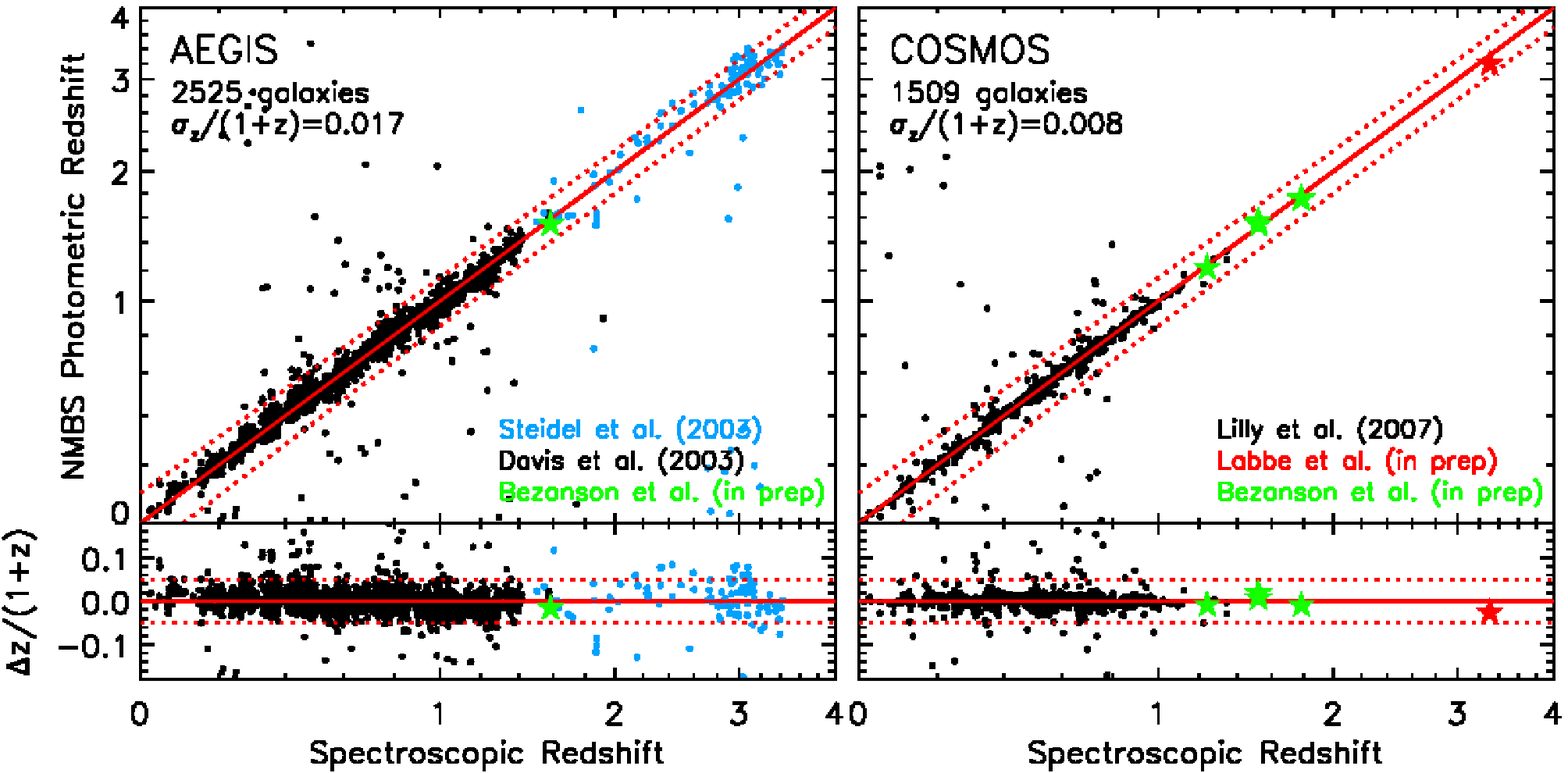}
\caption{Comparison between the NMBS photometric redshifts of the deblended catalog and
spectroscopic redshifts from the DEEP-2 sample
\citep{Davis03} and a smaller sample of high redshift LBGs \citep{Steidel03} (left panel), and
the $z$COSMOS survey \citep{Lilly07} (right panel).  A few preliminary spectroscopic redshifts
from Bezanson et al. ({\it in preparation}) and Labb\'{e} ({\it in preparation}) are
included.  For reference, the red dashed lines are for $z_{\mathrm{phot}}=z_{\mathrm{spec}}\pm0.05(1+z_{\mathrm{spec}})$.
The correspondence between the photometric and spectroscopic redshifts is generally very good.}
\label{fig:specz}
\end{figure*}

%=== Fig 14                                                                      
\begin{figure}[t!]
\leavevmode
\centering
\includegraphics[scale=0.45]{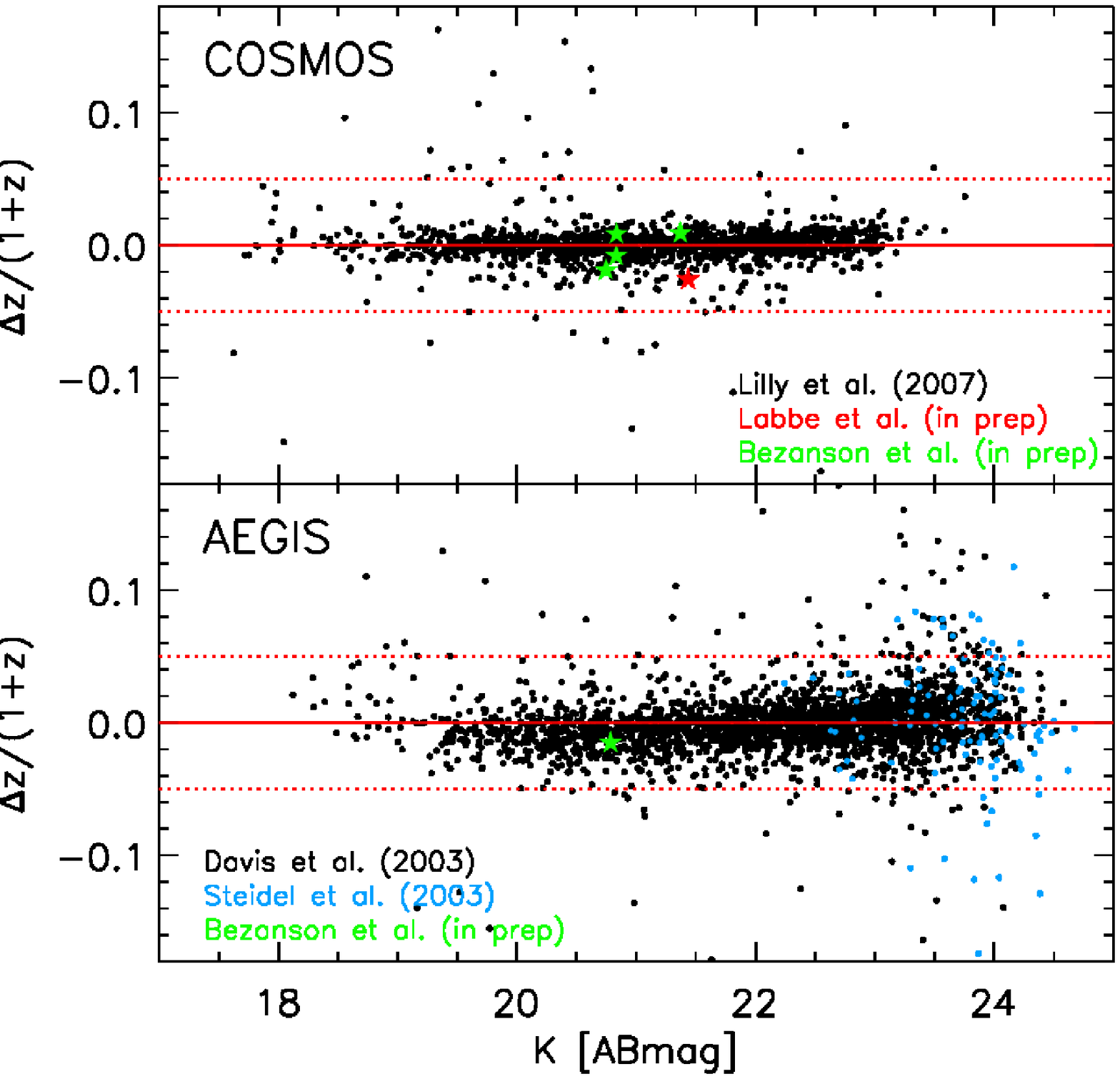}
\caption{The scatter in the photometric and spectroscopic redshifts of the deblended catalogs is relatively 
uniform as a function of the $K$-band magnitude, with a larger scatter amongst the faint, high redshift LBGs (blue).
The spectroscopic redshifts are from the $z$COSMOS survey \citep{Lilly07} (top panel) and the DEEP-2 
sample \citep{Davis03} with a small sample of high redshift LBGs \citep{Steidel03} (bottom panel).
We also include preliminary spectroscopic redshifts
from Bezanson et al. ({\it in preparation}) and Labb\'{e} ({\it in preparation}).  
The red dashed lines are for $z_{\mathrm{phot}}=z_{\mathrm{spec}}\pm0.05(1+z_{\mathrm{spec}})$.}
\label{fig:speczK}
\end{figure}

We have compared our $K$- and $K_{S}$-band flux measurements to the COSMOS survey 
catalog~\citep{Capak08,Ilbert09}\footnote{\url{http://irsa.ipac.caltech.edu/data/COSMOS/tables/redshift/}}.  The COSMOS catalog is $I$-selected 
and quotes fluxes within a 3$^{\prime\prime}$ aperture with an aperture correction provided to convert to AUTO fluxes
\citep[see][]{Capak07, Capak08}.  
The $K_{S}$-band data in the COSMOS catalog is derived from an 
earlier CFHT $K_S$ image taken during the 2004--2005 observing season, 
whereas we use the publicly available $K_S$ image with photometry from the 2005--2007 observing seasons
for this analysis.  As the COSMOS $K_S$ image used to make the publicly available COSMOS catalog has 
1.5$^{\prime\prime}$ seeing (compared to $\sim0.8^{\prime\prime}$ seeing in the image version used in the NMBS
catalogs), we smooth our $K_S$ image with a Gaussian kernel to 1.5$^{\prime\prime}$ and compare the fluxes 
between the catalogs.

Figure~\ref{fig:compareK} shows the difference in $K_{S}$-band magnitude between the smoothed NMBS $K_{S}$ image
and the magnitude quoted within a 3$^{\prime\prime}$ aperture in the COSMOS catalog (black points) for all objects.  
The results indicate that there are no significant systematic zero point offsets between the NMBS
and COSMOS catalogs.

We have also compared the GALEX and IRAC flux measurements to publicly available catalogs,
finding excellent agreement.  We exclude comparisons to catalogs where it is unclear how aperture corrections 
are applied as the interpretation is difficult, but note that there were not any large systematic offsets.  
The COSMOS GALEX NUV and FUV catalog fluxes within 3$^{\prime\prime}$ apertures 
agree within $\sim0.05$ magnitudes with the public catalogs available through the Multimission 
Archive\footnote{\url{http://archive.stsci.edu/prepds/cosmos/galexdatalist.html}}, 
The AEGIS NMBS IRAC photometry within 3.5$^{\prime\prime}$ apertures
has a median difference of $<0.03$ magnitudes for all channels when compared to the EGS 
catalog\footnote{\burl{http://www.cfa.harvard.edu/irac/egs/}}. Finally, we have also compared the MIPS total
fluxes in COSMOS to the sCOSMOS GO3 catalog, finding excellent agreement with a median offset of $<1\%$.  
In general, the photometry of the NMBS catalogs agrees well with all publicly available catalogs within the fields.
 
\section{Photometric Redshifts}
\label{sec:photoz}

Photometric redshifts and rest-frame colors were derived using the EAZY code\footnote{\url{http://www.astro.yale.edu/eazy/}}
\citep{Brammer08}, that fits linear combinations of seven templates to the broad- and medium-band
SEDs.  A $K$-band apparent magnitude prior and systematic errors due to template mismatch are taken into account.  
The optimized template set used for the EAZY photometric redshift code contains seven templates, which is large enough to 
span a broad range of galaxy colors, while minimizing color and redshift degeneracies.  The default template set is described 
in detail in \citet{Brammer08}; this set includes five templates generated based on the P\'{E}GASE 
population synthesis models \citep{Fioc99} and calibrated with 
synthetic photometry from semi-analytic models, as well as an additional young and dusty template added to compensate 
for the lack of extremely dusty galaxies in semi-analytic models.  We include an additional template for an old, red galaxy from the
\citet{Maraston05} models, with a \citet{Kroupa01} IMF and solar metallicity for a stellar population that has an age of 12.6 Gyrs.
The choice of template set is important for both the photometric redshifts and rest-frame colors \citep[see Appendix C of][for log M$_{\odot}>11$]{Whitaker10}.
We adopt {\tt z\_peak} as the photometric redshift (except where otherwise noted), which finds discrete peaks in the
redshift probability function and returns the peak with the largest integrated probability.

\begin{table}[t!]
  \caption{Zero Point Offsets}
  \centering
  \begin{threeparttable}
    \begin{tabular}{lcc}
      \hline
      \hline
      ~~~~~~~~~~~ & AEGIS\tnote{a} & COSMOS \\
      \hline
      $U$ & -0.22 & -0.23\\
      $B$ & - & -0.07\\
      $G$ & -0.01 & -0.02 \\
      $V$ & - & 0.12\\
      $R$ & -0.01 & -0.03\\
      $Rp$ & - & 0.03\\
      $I$ & 0.03 & 0.00\\ 
      $Ip$ & - & -0.34\\
      $Z$ & 0.05 & -0.01 \\
      $Zp$ & - & 0.11\\
      $J$ & -0.03 & -0.08\\
      $J_{1}$ & -0.00 & -0.01\\
      $J_{2}$ & -0.00 & -0.01\\
      $J_{3}$ & -0.01 & -0.02\\
      $H$ & -0.03 & -0.02\\
      $H_{1}$ & 0.01 & -0.01\\
      $H_{2}$ & -0.01 & -0.03\\
      $K$ & 0.00 & 0.00 \\
      $K_S$ & 0.04 & -0.05 \\
      $IA427$ & - & -0.11 \\
      $IA464$ & - & -0.16 \\
      $IA484$ & - & -0.11 \\
      $IA505$ & - & -0.02 \\
      $IA527$ & - & -0.06 \\
      $IA574$ & - & -0.12 \\
      $IA624$ & - & 0.04 \\
      $IA679$ & - & 0.24 \\
      $IA709$ & - & 0.15 \\
      $IA738$ & - & -0.04 \\
      $IA767$ & - & 0.07 \\
      $IA827$ & - & -0.12 \\
      \hline
    \end{tabular}
    \begin{tablenotes}
    \item[a] All zero point offsets are in magnitudes, defined as ZP$_{\mathrm{EAZY}}$ = ZP$_{\mathrm{nominal}}$ + offset
    \end{tablenotes}
  \end{threeparttable}
  \label{tab:zpoffset}
\end{table}

Uncertainties in the photometric zero points of the numerous photometric bands can lead to systematic 
shifts in the observed colors, and thus the redshift estimates, of the sample.  We compute offsets of the 
photometric zero points by fitting the EAZY templates to the full optical-NIR SEDs of objects with 
spectroscopic redshift measurements, with the fit redshift fixed to the spectroscopic value.  The offsets 
are computed iteratively by minimizing the de-redshifted fit residuals, which allows the 
separation of zero point errors from systematic effects associated with the choice of templates, as a 
given rest-frame wavelength is sampled by different observed bands at different redshifts.  
The zero point offsets are listed in Table~\ref{tab:zpoffset} and included in the catalog.  

While the majority of the bands require only small offsets ($\lesssim 0.02$ mag), some of the optical bands require 
offsets of as much as 0.3 mag \citep[see also Table 1 in][]{Ilbert09}.  
We have verified that this technique
reliably recovers artificial zero point errors of 0.02--0.3 mag added to the observed photometry.
The CFHTLS $u$-band zero point is uncertain and known to have issues; a direct comparison between the CFHT $u$-band
and SDSS suggests a difference of 0.1 mag~\citep{Erben09}.
The derived $u$-band zero point offset may be partly influenced by the choice of template set.
As a result, it is difficult to separate true zero point problems from template mismatch.
Furthermore, the $u$-band zero point is not well constrained as it is
not bracketed by other filters --- except the GALEX filters, which have their own known zero
point uncertainies up to $\sim0.3$ mag~\citep[e.g.,][]{Ilbert09}.

%=== Fig 15                                                                                               
\begin{figure}[t!]
\leavevmode
\centering
\includegraphics[scale=0.47]{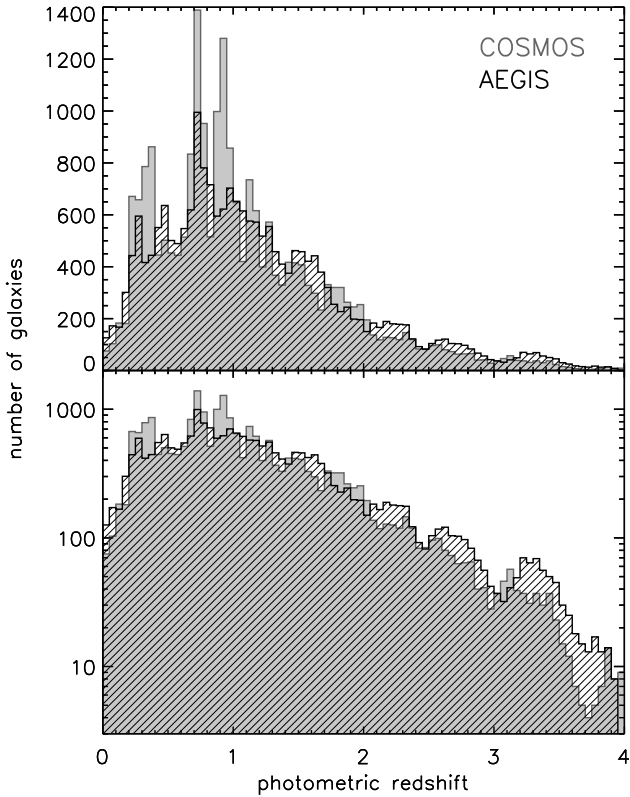}
\caption{The linear (top) and logarithmic (bottom)
photometric redshift distributions of AEGIS (black) and COSMOS (grey), using the EAZY {\tt z\_mc}
redshift~\citep[see definition from][]{Wittman09}. We use the standard selection of galaxies (use=1) with
$K_{\mathrm{tot}}$ greater than the 50\% completeness limits from the unmasked simulations (23.6 and
23.4 mag in AEGIS and COSMOS).
The spikes coincide with known overdensities.  There are 12,632 galaxies at $z>1.5$ shown here.}
\label{fig:zdistr}
\end{figure}

We find the photometric redshifts in COSMOS to be in excellent agreement with the magnitude-limited $I$-band 
$I_{\mathrm{AB}}<22.5$ spectroscopic sample of redshifts from the 
$z$COSMOS survey \citep{Lilly07},
with $\sigma_{z}$/(1+$z$)=0.008 for 1509 galaxies (right panel, Figure~\ref{fig:specz}), where $\sigma_{z}$/(1+$z$) is the
normalized median absolute deviation, or $\sigma_{\mathrm{NMAD}}=1.48\times\mathrm{MAD}$.  These values are quoted only for $z_\mathrm{spec}<1$, as there
are very few spectroscopic redshifts currently publicly available in the COSMOS field beyond this redshift.  We also
find excellent agreement between the photometric and spectroscopic redshifts for a larger sample
of 2525 objects at $z_\mathrm{spec}<1.5$ in AEGIS selected to a $R$-band limit of
$R_{\mathrm{AB}}<24.1$ from the DEEP2 survey \citep{Davis03} with
$\sigma_{z}$/(1+$z$)=0.017 (left panel, Figure~\ref{fig:specz}). Both fields have very few catastrophic failures, 
with only $\sim$5\% $>5\sigma$ outliers, defined such that 
$\left|z_{\mathrm{phot}}-z_{\mathrm{spec}}\right|$/(1+$z_{\mathrm{spec}}$) $<$ 5$\sigma_{\mathrm{NMAD}}$.
Spectroscopic redshifts are available for 117 Lyman break galaxies (LBGs) at $z\sim$3 within the 
AEGIS field from \citet{Steidel03},
for which we find $\sigma_{z}$/(1+$z$)=0.044, with 10\% $>5\sigma$ outliers.  These relatively high photometric
redshift uncertainties for LBGs are well known~\citep{Reddy08} and caused by the fact that
LBGs tend to be faint in the rest-frame optical (observer's NIR) and their spectra have relatively
weak Balmer/4000\AA\ breaks.  We also note that there is excellent agreement between the NIR
medium-band photometric redshifts and the 
Gemini/GNIRS redshifts from \citet{Kriek06}, with a biweight scatter in 
$\left|z_{\mathrm{phot}}-z_{\mathrm{spec}}\right|$/($1+z_{\mathrm{spec}}$) of only 0.010, albeit this is only for 
four galaxies \citep[see][]{vanDokkum09a}.  The photometric redshift accuracies range from 
$\sigma_{z}/(1+z)\sim1$\% for galaxies with stronger Balmer/4000\AA\ breaks to 
$\sigma_{z}/(1+z)\sim5$\% for galaxies with less defined breaks from the spectroscopic redshift sample.  
Recently, \citet{Kriek11} find that the photometric redshift errors must be $\lesssim2$\% given the shape of the
observed H$\alpha$ emission line, as built from composite SEDs of the medium-band photometry itself.

In an effort to fill in the void of spectroscopic redshifts for $K$-selected galaxies above $z\sim1$, 
follow-up spectroscopy
for several massive galaxies at $z\sim1.5$--3.5 have been persued.  Preliminary spectroscopic redshifts
are in excellent agreement with the NMBS photometric redshifts (see Bezanson et al. {\it in preparation}
and Labb\'{e} et al. {\it in preparation}).  We include the currently available preliminary 
spectroscopic redshifts in the right panel of Figure~\ref{fig:specz}.

%=== Fig 16                                                                                                          
\begin{figure}[t!]
\leavevmode
\centering
\includegraphics[scale=0.45]{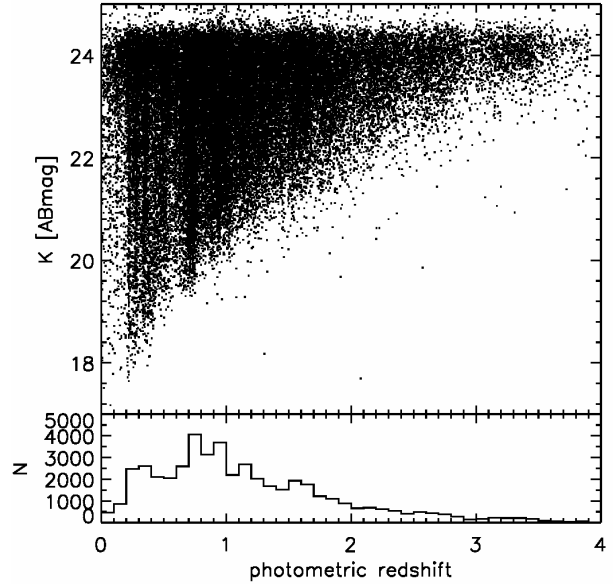}
\caption{The distribution of apparent $K$-band magnitude as a function of the photometric redshift
({\tt z\_peak}) for all galaxies in the NMBS fields (top) and the number of galaxies as a
function of their photometric redshift (bottom).}
\label{fig:Kmagzpeak}
\end{figure}

%=== Fig 17                                                                                                      
\begin{figure*}[t!]
\leavevmode
\centering
\includegraphics[scale=0.65]{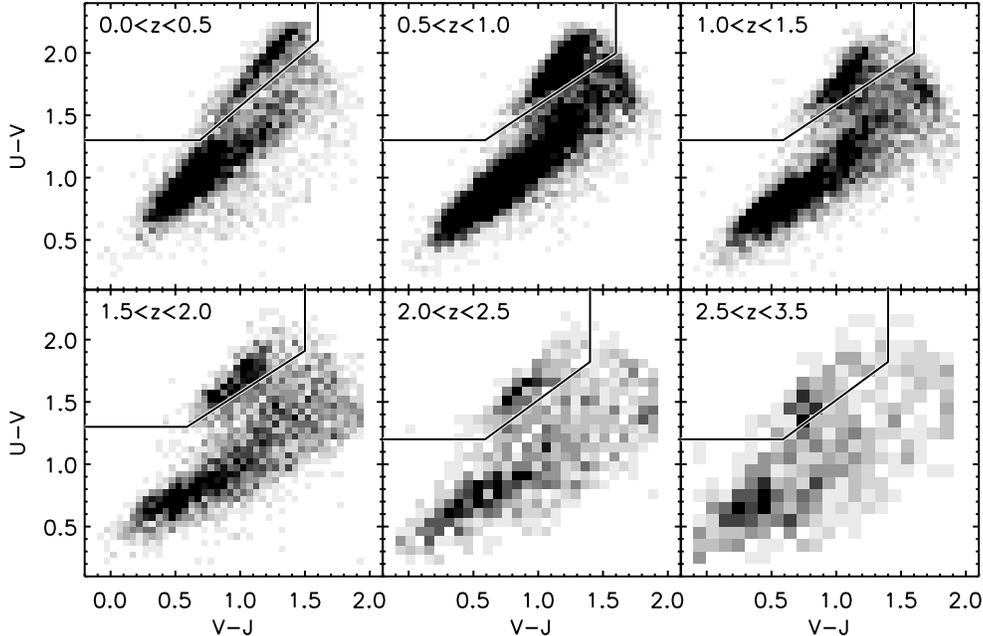}
\caption{The rest-frame $UVJ$ diagram for NMBS galaxies with $\mathrm{S/N}>8$ in the $K$-band out to a redshift of 3.\
5.
The grey scale represents the density of points, with the lines indicating the separation between
quiescent ``red sequence'' galaxies and star forming galaxies
(both blue and red).  The quiescent galaxies can be traced to the highest redshift interval $2.5<z<3.5$.
Apparently, the first quiescent galaxies stopped forming stars by that redshift \citep[see also][]{Marchesini10}.}
\label{fig:UVJ}
\end{figure*}

In Figure~\ref{fig:speczK}, we see that the scatter in the photometric-spectroscopic redshift comparison is relatively
uniform as a function of the $K$-band magnitude.  There is a larger scatter amongst the high redshift LBGs (blue),
which is not surprising given the spectral shapes and faint $K$-band magnitudes of these galaxies.

The photometric redshift distributions of both fields are shown in Figure~\ref{fig:zdistr}, using 
the EAZY {\tt z\_mc} redshift.  
This redshift value is drawn randomly from the redshift probability distribution, where the 
distribution of these redshifts for a given sample of objects very closely follows the 
summed probability distribution of those same objects~\citep{Wittman09}.  The fields
have independent spikes in redshift space, indicating the presence of known 
overdensities~\citep[see, e.g.,][]{Olsen07}.
The overdensities in COSMOS at $z\lesssim1$ are more prominent due to the addition of the 
twelve medium-band optical filters and may also be the partly due to the relative 
variance in the size of the structures from field to field.
Additionally, we show the distribution of apparent $K$-band magnitude as a function of photometric 
redshift (now {\tt z\_peak}, as throughout the rest of the paper) of all galaxies in the NMBS 
fields in Figure~\ref{fig:Kmagzpeak}.  

\section{Evidence for Quiescent Galaxies to $z\sim3$}
\label{sec:quiescent}

%=== Fig 18                                                                                          
\begin{figure*}[t!]
\leavevmode
\centering
\includegraphics[scale=0.75]{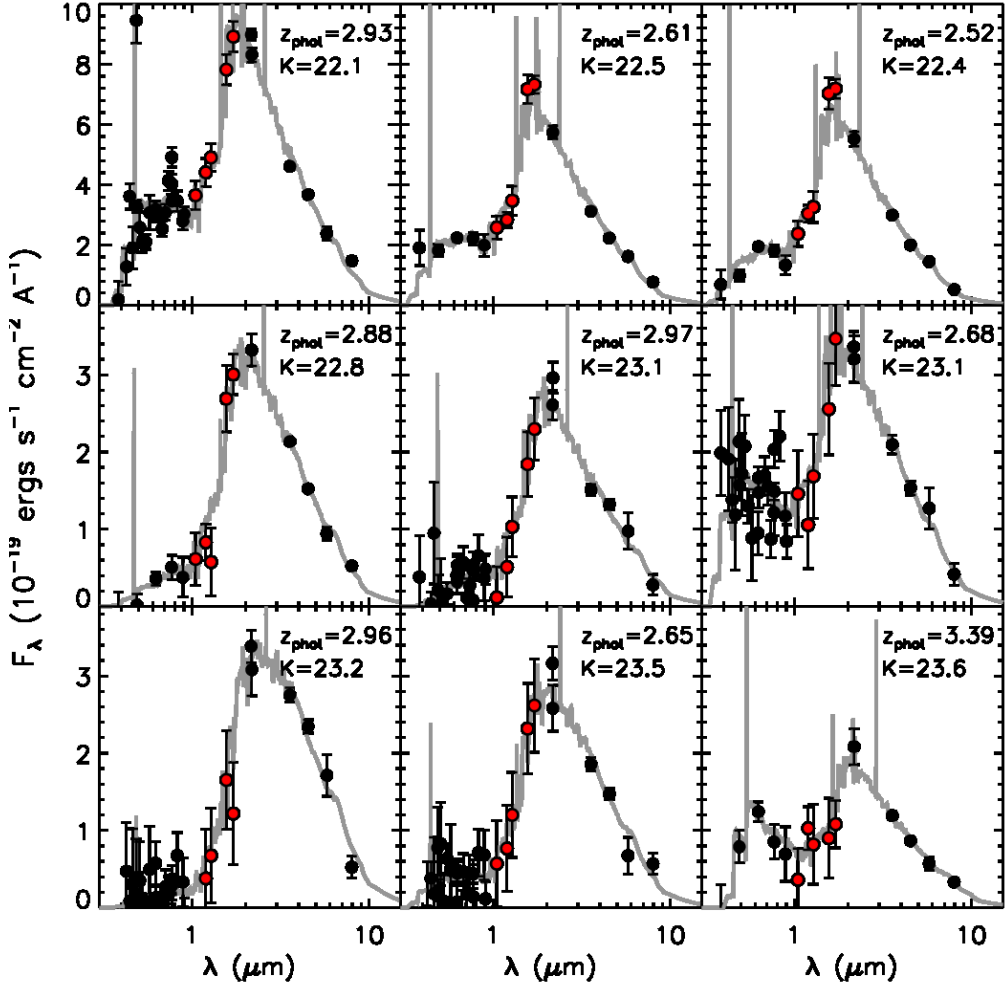}
\caption{Example quiescent SEDs selected based on their $U$--$V$ and $V$--$J$ rest-frame colors
from the highest redshift bin in Figure~\ref{fig:UVJ}, sorted from bright to faint in $K$-band magnitude.
The best-fit EAZY template is shown in grey and the medium-band filters in red.
All of these galaxies show a prominent Balmer/4000\AA\ break. }
\label{fig:UVJexample}
\end{figure*}

Galaxies within the local universe lie in two distinct classes: actively star-forming galaxies
and ``quiescent'' galaxies with evolved stellar populations and little ongoing star-formation.
These populations show strong bimodal behavior in their colors~\citep{Kauffmann03,Baldry04}, where the red,
quiescent galaxies have a more luminous characteristic magnitude and a shallower faint-end slope of
the luminosity and mass functions compared to those of blue galaxies.

Quiescent galaxies fall into a distinct ``red sequence'' in color-magnitude diagrams at low redshift.  However,
at high redshifts dusty, star-forming galaxies can have similarly red $U$--$V$ colors, ``contaminating'' 
the red sequence.  Including the $V$--$J$ rest-frame color can help separate out the dusty, star-forming
galaxies~\citep{Labbe07,Wuyts07,Williams09,Ilbert09}.  
At fixed $U$--$V$, passively evolving galaxies have blue $V$--$J$ colors, whereas dusty galaxies 
have red $V$--$J$ colors.
The location of the red clump of quiescent galaxies in the $UVJ$ plane shifts towards bluer $U$--$V$ colors
towards higher redshift, while the $V$--$J$ colors remain essentially unchanged.
This change in rest-frame color with redshift reflects the intrinsic passive evolution of the stellar population.
However, the uncertainties in the rest-frame colors will also increase with redshift, potentially
washing out any intrinsic bimodality.  These two effects have made detections of the bimodality 
difficult at high redshift. 

This bimodal color sequence has been shown to persist out to $z\sim2$ both 
photometrically~\citep[e.g.,][]{Labbe05, Daddi05, Taylor09a,
Williams09,Whitaker10} and spectroscopically~\citep[e.g.,][]{Giallongo05,Franzetti07,Cassata08,Kriek08}.  
A population of massive, quiescent galaxies at $z\sim2$
is surprising.  The ages of the stellar populations of these galaxies are a significant fraction of
the age of the universe, which was only a few Gyr at these redshifts.  The existence of 
massive galaxies with strongly suppressed star formation at such high redshifts provides useful 
constraints on models of galaxy formation~\citep[e.g.,][]{Baugh06}.
It is unclear at what point in cosmic time that this bimodal color sequence emerges, due
to limitations governed by the accuracies of the photometric redshifts and rest-frame colors.

The accurate redshifts and rest-frame colors of the NMBS are well suited for studies 
of the evolution of the bimodal color distribution of galaxies. In \citet{Brammer09}, we selected 25,000 
galaxies from the NMBS showing that the dust-corrected rest-frame $U$--$V$ color distribution is bimodal out to
$z\sim2.5$.
Additionally, \citet{Whitaker10} measured an increase in the intrinsic scatter of the rest-frame $U$--$V$ 
colors of massive, quiescent galaxies in the NMBS out to $z\sim2$.  This scatter in color arises from the 
spread in ages of the quiescent galaxies, where we see both quiescent red, old galaxies and 
quiescent blue, younger (post-starburst) galaxies toward higher redshift.  Both of these studies were based on 
NMBS catalogs from the 2008A and 2008B data alone, whereas here we are using a deeper catalog 
containing the entire survey data set.  Also using the latest NMBS catalog,~\citet{Marchesini10} showed that 
the most massive galaxies ($>2.5\times10^{11}$ M$_{\odot}$) at $3<z<4$ are typically red and faint 
in the observer's optical with quenched star-formation, extending the 
evidence for quiescent galaxies beyond $z\sim3$ with a sample of 14 very massive galaxies.

Rest-frame $U-V$ and $V-J$ colors are derived by integrating the redshifted rest-frame filter bandpasses
from the best-fit EAZY template.  We use the $UBV$ bandpasses defined by \citet{Maiz06} and the 2MASS $J$ bandpass.
This method for determining rest-frame fluxes is similar to that used by the COMBO-17 survey described by
\citep{Wolf03}.  We find that in most cases this method provides very similar results to the algorithm described
by~\citet{vanDokkum96} and~\citet{Taylor09a}, though the direct template fluxes are
more robust when narrower, closely-spaced filters are used, as is the case here.
A more detailed explanation of the rest-frame color algorithm will be provided by \citet{Brammer11}.
We note that the observed $u$-band is 
used in determining the rest-frame $U$--$V$ colors for low redshift galaxies ($z\lesssim0.4$).
In order to limit biases, we do not use the zero point offset which gives
the best template fits in the rest-frame color determinations due to the large uncertainties.

Figure~\ref{fig:UVJ} shows the rest-frame $U$--$V$ versus $V$--$J$ diagram for the NMBS galaxies, selecting galaxies
with a S/N$>8$ in the $K$-band.  For the first time, we show the
bimodal distribution of star-forming and quiescent galaxies based on their rest-frame colors out to a
remarkably high redshift of $z\sim3$.  A pronounced ``quiescent clump'' is visible at red $U$--$V$ colors above the star-forming
sequence out to $z=2.5$, with some evidence of a bimodality at $2.5<z<3.5$.  
This clump appears to shift to bluer colors at higher redshift.
We note that the shape of the red clump of quiescent galaxies at $0<z<0.5$
extends to bluer $U$--$V$ colors, whereas the analogs of these galaxies appear to absent at all other redshifts.
This effect may be caused by low mass quiescent galaxies with lower metallicities that are below our $K$-band detection
limit at higher redshifts.  
We also note that the $U$--$V$, $V$--$J$ plane would be populated differently in an 
optically-selected survey: there would be more faint, blue galaxies that fall in the lower left corner.
The incompleteness of this flux-limited sample leads to an apparent increase in the number of 
quiescent galaxies relative to blue, star-forming galaxies at the highest redshifts.  Using a mass-selected sample
of quiescent galaxies, \citet{Whitaker10} have shown that the fraction of quiescent galaxies falls below 40\% at $z>2$.

As stellar population parameters are dependent on the 
stellar population synthesis models~\citep[e.g., see][]{Wuyts07,Muzzin09,Conroy09, Conroy10}, we adopt
the model-independent selection criteria of quiescent galaxies of \citet{Williams09} based on the rest-frame $U$--$V$ and $V$--$J$ colors.
We have modified the selection limits of \citet{Williams09} at $z>1$, with the following adopted diagonal selection criteria,

\begin{eqnarray}
(U-V)>0.88\times(V-J)+0.69~~~~~~[z<&0.5&]\nonumber\\
(U-V)>0.88\times(V-J)+0.59~~~~~~[z>&0.5&]
\end{eqnarray}

Additionally, we modify the limits in $U$--$V$ and $V$--$J$ such that,
\begin{eqnarray}
(U-V)>1.3,~(V-J)<1.6 ~~~~~~[0.0<z<&1.5&]~~~~~ \nonumber\\
(U-V)>1.3,~(V-J)<1.5 ~~~~~~[1.5<z<&2.0&]~~~~~ \\
(U-V)>1.2,~(V-J)<1.4 ~~~~~~[2.0<z<&3.5&]~~~~~ \nonumber
\end{eqnarray}

\noindent where these constraints on the $U$--$V$ and $V$--$J$ colors prevent contamination from unobscured and 
dusty star–forming galaxies, respectively. 
The clump of quiescent galaxies is separated by only
$\sim0.3$ mag in $U$--$V$ color from the blue, star-forming sequence at $z\sim3$.  
We note that this offset is larger than the typical uncertainties in the rest-frame colors at $z\sim3$ 
of $<0.1$ mag.  We speculate that if we were able to 
resolve enough galaxies at $z\sim3.5$--4, we might begin to see the epoch where this color bimodality breaks down.

In Figure~\ref{fig:UVJexample}, we demonstrate the quality of the SEDs of quiescent galaxies in the highest 
redshift bin of Figure~\ref{fig:UVJ}.  The $K$-band S/N of these $z\sim3$ 
galaxies ranges from S/N$\sim35$ for the brightest galaxy 
($K\sim22$ mag) down to S/N$\sim9$ for the faintest galaxies ($K\sim24$ mag).  The majority of the galaxies
shown in Figure~\ref{fig:UVJexample} have little to no rest-frame UV emission, indicating little ongoing star
formation.  We note that the rest-frame optical and NIR colors are dominated by an old stellar population, and that a 
comparatively low level of star formation is possible even on the ``quiescent'' side of the bimodal population.

\section{Improvements enabled by the NIR Medium-band Filters}
\label{sec:benefits}

%=== Fig 19
\begin{figure}[b!]
\leavevmode
\centering
\includegraphics[scale=0.45]{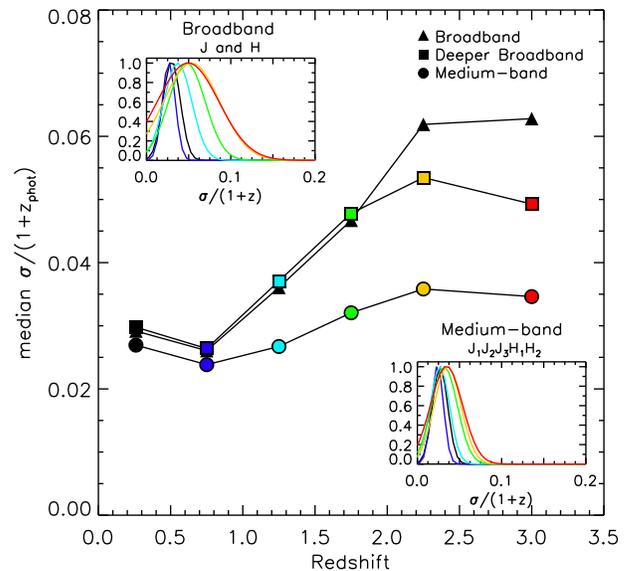}
\caption{The median 68\% confidence interval, $\langle\sigma/(1+z)\rangle$, of the photometric redshifts of
quiescent galaxies selected by the $UVJ$ rest-frame color selection technique.  The confidence intervals
are calculated for medium-band catalogs and broadband catalogs that
replace the $J_{1}$, $J_{2}$, $J_{3}$, $H_{1}$ and $H_{2}$ medium-band filters with $J$ and $H$
filters.}
\label{fig:conint}
\end{figure}

%=== Fig 20
\begin{figure}[t!]
\leavevmode
\centering
\includegraphics[scale=0.65]{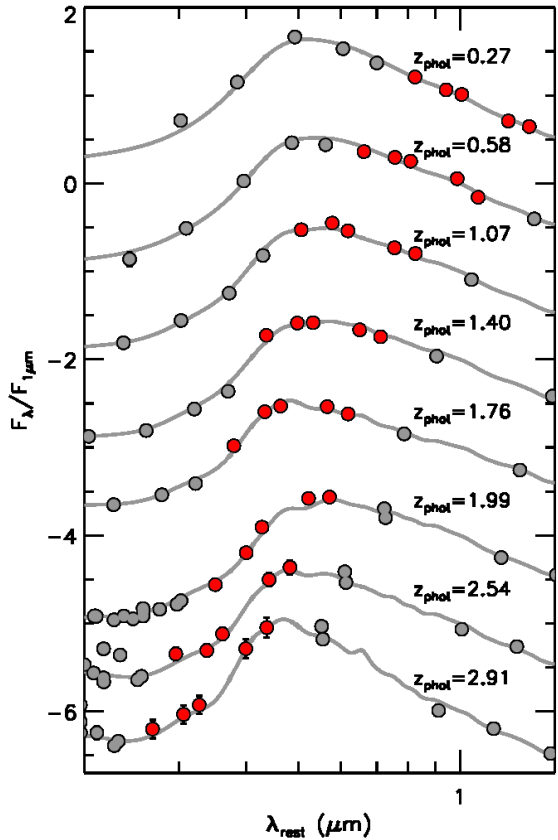}
\caption{Examples of quiescent galaxies ranging from $z=0.27$ to $z=2.91$, showing the rest-frame wavelength
range the NIR medium-bandwidth filters span (red) as a function of redshift.  The best-fit EAZY models are
smoothed with a Gaussian with a width of $0.15\mu$m/$(1+z)$, the resolution of the medium-band filters. The
medium-band filters begin to sample the Balmer/4000\AA\ region at $z>1.5$.}
\label{fig:quiescentexample}
\end{figure}

Throughout this paper, we have emphasized the superior quality of the NMBS photometric redshifts
and consequently, rest-frame colors.  Here, we provide a more comprehensive look at the improvements provided
by medium-band filters compared to the standard broadband filters commonly used in high redshift
photometric surveys.  

In \S~\ref{sec:quiescent}, we isolated a sample of quiescent galaxies out to $z=3.5$ from the
NMBS using the rest-frame $U$--$V$ and $V$--$J$ colors.  Although these galaxies span a large range of S/N,
their spectral shapes will unambiguously contain strong Balmer/4000\AA\ breaks.  This sample of 
galaxies is therefore ideal to characterize the power of the NIR medium-band filters relative to the
canonical broadband filters.  

We have generated broadband catalogs in both fields with the same depth as the medium-band images, as well as a 
broadband catalogs with the deeper WIRDs $J$ and $H$ photometry.  The WIRDS $J$- and $H$-band images have depths that are
1.2 and 2.3 mag deeper than the NMBS medium-band images in AEGIS (0.1 and 0.6 mag deeper in COSMOS), enabling us to 
test how depth affects the accuracies of the photometric redshifts.  The images with the same depth as the NMBS
survey were generated by combining the $J_{2}$ and $J_{3}$ medium-band images 
to create a $J$-band image ($J=(J_{2}+J_{3})/2$), and the $H_{1}$ and $H_{2}$ medium-band images 
for an $H$-band image ($H=(H_{1}+H_{2})/2$) for both fields.  The broadband catalogs contain FUV/NUV, 
$ugrizJHK$ and IRAC photometry in both fields and we compare these to medium-band catalogs with 
FUV/NUV, $ugrizJ_1J_2J_3H_1H_2K$ and IRAC photometry.  

The photometric redshifts are fit for the medium-band and broadband catalogs
using the same parameter settings 
and templates as described in \S~\ref{sec:photoz}.  Figure~\ref{fig:conint} compares the 
median 68\% confidence interval, $\langle\sigma/(1+z)\rangle$, of the photometric redshifts for
all quiescent galaxies in the medium-band and broadband catalogs.  
The confidence intervals of photometric redshifts are appropriate given the true scatter in 
Figure~\ref{fig:specz} (at least for $z<1$).  It is therefore reasonable to believe them, and hence 
a comparison of confidence intervals for the larger sample is meaningful.
We limit the sample to a completeness level of $\gtrsim$90\%, with $K_{\mathrm{tot}}<22.2$ mag. 
We note that the COSMOS field includes roughly twice as many optical filters than the AEGIS field,
including the twelve {\it Subaru} optical medium-band filters.  The additional deep optical broad- and medium-bands 
improve the confidence intervals of the COSMOS NMBS photometric redshifts at $z<1$ by a factor of two, 
with little effect at $z>1$.
For simplicity, we restrict this analysis to only those filters included in both fields 
where we can strictly test the 
differences between a standard broadband catalog and the addition of the NIR medium-band filters.

%=== Fig 21
\begin{figure}[t!]
\leavevmode
\centering
\includegraphics[scale=0.45]{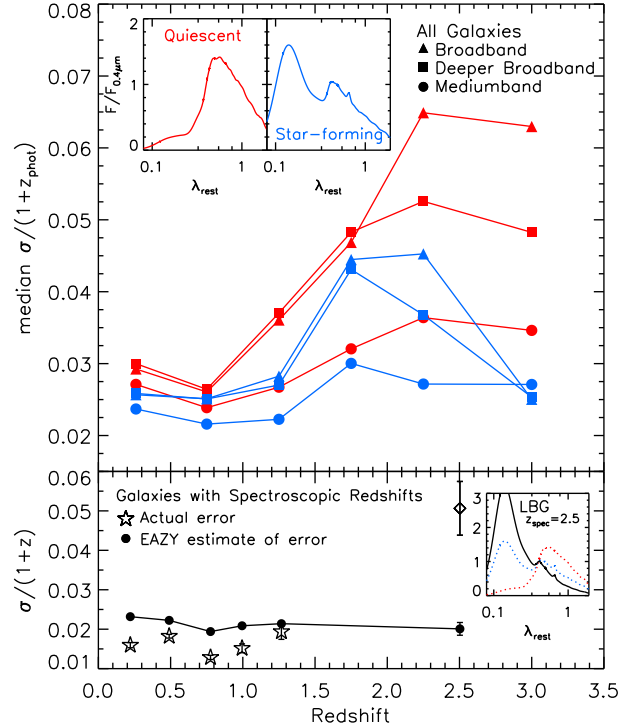}
\caption{The median 68\% confidence interval of the photometric redshifts of quiescent (red) and star-forming
(blue) galaxies selected by the $UVJ$ rest-frame color selection technique.  The confidence intervals are calculated
for both the medium-band and broadband catalogs.  The top right inset panels show the difference between
the average best-fit EAZY templates of the quiescent (left) and star-forming (right) galaxies, smoothed with a Gaussian
to the resolution of the medium-band filters.  The bottom panel shows the median confidence interval of only those 
galaxies with spectroscopic redshifts compared to the actual errors (stars) as determined from the photometric-spectroscopic
comparison in \S\ref{sec:photoz}, with the LBGs indicated separately as a diamond.  The best-fit EAZY template for the sample of LBGs at $z_{\mathrm{spec}}=2.5$ is bluer
than the average star-forming galaxy, with a weak Balmer/4000\AA\ break (see bottom right inset panel).  The error for both quiescent and star-forming galaxies is up to a factor of two times smaller when using medium-band NIR data as compared to deep broadband data.}
\label{fig:conint2}
\end{figure}

The median 1$\sigma$ confidence intervals of the broadband photometric redshifts of quiescent galaxies are 
up to a factor of two times greater than those calculated with the NIR medium-band filters. 
We note that the formal errors on the photometric redshifts will scale with the S/N of the detected 
objects, where the median 1$\sigma$ confidence intervals will be smaller for brighter samples within 
the NMBS catalogs.  For example, the typical 1$\sigma$ confidence interval for a galaxy with 
$K_{\mathrm{tot}}$=18--20 is 0.02, whereas this value increases to $\sim0.04$ for $K_{\mathrm{tot}}$=22.
The distribution of $\sigma/(1+z)$ for the quiescent sample within each redshift bin
is roughly Gaussian; the inset panels in Figure~\ref{fig:conint} show the best-fit Gaussian distributions.
The width of the distributions range from 0.01--0.02 for the medium-band catalogs,
compared to a factor of two increase for the broadband catalogs.  

The difference between the medium-band and broadband catalogs becomes more pronounced for those quiescent galaxies
with $z>1.5$, where the NIR medium-band filters start to sample the rest-frame
Balmer/4000\AA\ break region.  Figure~\ref{fig:quiescentexample} shows examples of quiescent galaxies ranging 
from $z=0.27$ to $z=2.91$ with the best-fit EAZY models smoothed with a Gaussian of width $0.15\mu$m/$(1+z)$, 
the width of the medium-band filters.  The rest-frame wavelength range that is sampled by the NIR medium-bandwidth filters 
(shown in red) shifts to shorter wavelengths towards higher redshift, 
beginning to sample the Balmer/4000\AA\ region at $z>1.5$.

Figure~\ref{fig:conint2} compares the median confidence intervals for the photometric redshifts of quiescent (red) 
and star-forming (blue) galaxies in the medium-band and broadband catalogs, as selected from the 
$UVJ$ diagram in Figure~\ref{fig:UVJ}.  The difference between the broadband catalog and the addition
of the NIR medium-band filters is most important for quiescent galaxies at $z>1.5$.  
Deeper broadband photometry (squares in Figure~\ref{fig:conint2}) improves the errors at $z>1.5$, but only to a certain degree.  The formal errors will decrease as the error bars decrease with deeper photometry
but they seem to reach a floor around $\sigma/(1+z)\sim0.05$,
at which point increased spectral resolution is vital.

Quiescent and star-forming
galaxies have different spectral shapes: the only prominent spectral feature in the spectrum of the quiescent 
galaxies is the Balmer/4000\AA\ break whereas the star-forming galaxies typically have rest-frame UV emission
with strong Lyman breaks.  The top inset panels in Figure~\ref{fig:conint2} show the spectral shape of the average 
best-fit EAZY template for the quiescent and star-forming galaxies, smoothed with a Gaussian to the resolution of the
medium-band filters.  We note that the decrease in the width of the confidence
interval at $z=3$ for the blue, star-forming galaxies for the broadband catalog is the result of the Lyman break 
shifting into the optical filter passbands.  

The median confidence interval for the blue, star-forming galaxies are 
up to a factor of $\sim1.5$ smaller than for the red, quiescent galaxies for the catalog that includes
the medium-band filters.  The medium-band filters provide more similar photometric redshift uncertainties for the 
quiescent and star-forming samples (i.e. they reduce the huge divergence seen at $z>2$ for standard broadband filters), 
which can be important for clustering analyses. 
We note that the average blue, star-forming galaxy still has a well-developed 
Balmer/4000\AA\ break, which is why the medium-band filters help even for those galaxies.  
Additionally, the deep optical broadband data will help constrain 
the photometric redshifts for those galaxies with rest-frame UV emission.  

\citet{vanDokkum06} showed that the ``median galaxy'' at $2<z<3$ with mass $>10^{11}$ M$_{\odot}$ 
is faint in the observer's optical, red in the 
observed NIR and has a rest-frame UV spectrum that is relatively flat in $F_{\lambda}$.  Therefore, the typical massive
high redshift galaxy can either have a spectral shape similar to quiescent galaxies or that of a red, dusty galaxy.  
For both cases, Figure~\ref{fig:conint2} demonstrates that 
the NIR medium-bandwidth filters are necessary to improve the confidence intervals of the photometric redshifts
of the typical massive high redshift galaxy.  In fact, the photometric redshifts are more accurate when including the NIR medium-band
filters regardless of the spectral shape.

In addition to comparing photometric redshift confidence intervals, we can also look at how the fraction 
of catastrophic outliers and the overall accuracy of those galaxies with spectroscopic redshifts changes 
in both fields when using the broadband catalogs versus the medium-band catalogs.  
We find that the fraction of catastrophic outliers increases by a factor of two in COSMOS when using
only broadband filters with similar depth to the NMBS survey.  By increasing the depth of the $J$- and $H$-band images
by 1-2 mag, as is the case with the AEGIS WIRDS data, the fraction of catastrophic outliers is the same as that 
resulting from the medium-band photometry.  
The scatter, $\sigma_{\mathrm{z}}/(1+z)$, increases by 15\% and 30\% in COSMOS and AEGIS for the deeper broadband catalogs,
consistent with the EAZY estimate of the errors at these redshifts (see Figure~\ref{fig:conint2}).  

We note that the EAZY confidence intervals are not well calibrated.
To test how representative the 68\% confidence intervals of the photometric redshifts are of the actual errors,
we consider the sub-sample of galaxies with spectroscopic redshifts.  In the bottom panel of Figure~\ref{fig:conint2},
we compare the actual errors from the photometric-spectroscopic redshift comparision in \S\ref{sec:photoz} to the median
confidence intervals.  At $z<1.5$, the EAZY error tends to
overestimate the actual error by $\sim20$--40\%.  On the other hand, the sample of LBGs at $z>1.5$ have actual errors that are almost
a factor of three times larger than the EAZY errors (diamond in bottom panel of Figure~\ref{fig:conint2}).  
We note that the LBGs have spectral
shapes different from the majority of NMBS galaxies (see bottom right inset panel in
Figure~\ref{fig:conint2}), with weak Balmer/4000\AA\ breaks and very faint $K$-band magnitudes.  The median $K$-band
magnitude of the sample of LBGs with spectroscopic redshifts is 23.9 mag, compared to a median $K$-band magnitude of
22.2 mag for the sample of galaxies with $z_{\mathrm{spec}}<1.5$ (see also Figure~\ref{fig:speczK}).
For brighter $K$-band magnitude limits at $z>1.5$, we currently have very limited spectroscopic information (see \S\ref{sec:photoz}).

\section{Summary}
\label{sec:summary}

We present the relatively deep and wide medium-band NIR imaging of the NMBS survey, 
consisting of two $\sim0.21$ deg$^{2}$ fields within the AEGIS and COSMOS surveys.  
The observations were carried out as part of a NOAO Survey Program on the Mayall 4m telescope on Kitt Peak,
using the NEWFIRM camera with five NIR medium bandwidth filters.
The full details of the observations and data reduction are described in \S\ref{sec:observations} and
\S\ref{sec:reduction}, respectively, including various
optimizations to the images to improve the image quality.  The astrometry is accurate within
$\lesssim0.1^{\prime\prime}$ (0.3 pixels).  The final combined images are constructed from 
individually registered, distortion-corrected, weighted averages of all frames; the flatness of the background 
and excellent quality of the data is readily apparent in Figures~\ref{fig:cosmosK} and~\ref{fig:aegisK}.
The reduced NMBS images and weight maps are publicly available through the NOAO archive and through
the NMBS site.  

We combined the NMBS data with UV (GALEX), visible (CFHT and {\it Subaru}), NIR (CFHT) and 
mid-IR ({\it Spitzer}/IRAC) data to produce a public $K$-selected photometric catalog,
which we make available to the community\footnote{\url{www.astro.yale.edu/nmbs}}.  
As about 10\% of the objects detected by SExtractor are clearly blended objects (for example, see 
Figure~\ref{fig:deblendsed}), we use custom scripts to deblend the objects using a higher resolution $K$-band
image.  The deblended catalog is included in addition to the original list of objects identified using SExtractor. 
With the NMBS catalogs, we summarize our main findings as follows:

\begin{enumerate}

\item The NMBS catalogs contain $\sim13,000$ galaxies at $z>1.5$ with accurate photometric
redshifts and rest-frame colors.  We demonstrate the excellent quality of the NMBS SEDs
throughout this paper~\citep[see also,][]{vanDokkum09a, Brammer09,vanDokkum10, Whitaker10, 
Kriek10, Marchesini10,Kriek11,Brammer11}.

\item We find excellent agreement with available spectroscopic
redshifts, with $\sigma_{z}/(1+z)\sim0.008$ for 1509 galaxies in COSMOS and
$\sigma_{z}/(1+z)\sim0.017$ for 2525 galaxies in AEGIS.  It should be noted that the currently available spectroscopic samples
are heavily weighted toward blue, low redshift galaxies.  Several programs for follow up spectroscopy 
of NMBS targets at $z>1.5$ are currently underway.

\item We show evidence of a clear bimodal color distribution between quiescent and star-forming galaxies persisting
to $z\sim3$, a higher redshift than has been probed before.

\item The median 68\% confidence intervals of the photometric redshifts for both quiescent and star-forming galaxies 
is up to a factor of two times smaller when comparing medium-band catalogs to broadband catalogs
due to the increased resolution of the five NIR medium-band filters.

\end{enumerate}

Follow up programs will continue to add to the legacy of the survey, as our understanding of these high redshift galaxies grows and evolves.
Calibrations of the medium-band photometric redshifts at $1<z<3.5$ will be possible with the upcoming 
HST Treasury program, 3D--HST.  This NIR spectroscopic survey will
partially overlap with the NMBS fields.

\begin{acknowledgements}
We thank the anonymous referee for useful comments and a
careful reading of the paper.
This paper is based partly on observations obtained with MegaPrime/MegaCam,
a joint project of CFHT and CEA/DAPNIA, and WIRCam, a joint project of 
CFHT, Taiwan, Korea, Canada and France, at the CFHT, which is operated by the National Research Council (NRC)
of Canada, the Institut National des Science de l'Univers of the Centre
National de la Recherche Scientifique (CNRS) of France, and the University
of Hawaii.  This work is based in part on data products produced at TERAPIX, the WIRDS 
consortium, and the Canadian Astronomy Data Centre. 
We thank H. Hildebrandt for providing the CARS-reduced CFHT-LS
images.  This study also makes use of data from AEGIS, a multiwavelength 
sky survey conducted with the Chandra, GALEX, Hubble, Keck, CFHT, MMT, 
Subaru, Palomar, Spitzer, VLA, and other telescopes and supported in part 
by the NSF, NASA, and the STFC.
Support from NSF grant AST-0807974 and NASA grant NNX11AB08G is gratefully acknowledged.
\end{acknowledgements}

\facility{\emph{facilities}: Mayall (NEWFIRM)}

\addcontentsline{toc}{chapter}{\numberline {}{\sc References}}
\bibliography{master}

\end{document}